%% file: SnowmassWhitePaper.tex
\documentclass[11pt,usetikz]{atlasnote}
\graphicspath{{figures/}}
\usepackage{subfig}
\usepackage{atlasphysics}
\usepackage{mathrsfs}
\usepackage{hyperref}
\usepackage{slashed}
\usepackage{multirow}

\skipbeforetitle{100pt}

\title{Physics at a High-Luminosity LHC with ATLAS}

\author{The ATLAS Collaboration}

\date{July 26, 2013 \\ Minor Revision: July 31, 2013}
\atlasnote{ATL-PHYS-PUB-2013-007}

\journal{the Snowmass Community Planning Study \\ \ Revised references with respect to the version of July 26, 2013}
\subprepend{Submitted as input to}

\abstracttext{
  The physics accessible at the high-luminosity phase of the LHC extends well
  beyond that of the earlier LHC program. Selected topics, spanning
  from Higgs boson studies to new particle searches and rare top quark decays, are
  presented in this document.  They illustrate the substantially enhanced
  physics reach with an increased integrated luminosity of $3000 \, \ifb$, and
  motivate the planned upgrades of the LHC machine and ATLAS detector.
}

\begin{document}

\input{coverletter}

\newpage
\input{introduction}

\input{detectors}

\input{higgs}

\input{vectorbosons}

\input{susy}

\input{exotics}

\input{top}

\input{conclusions}

\bibliographystyle{atlasBibStyleWoTitle}
\bibliography{SnowmassWhitePaper}

\end{document}

%% file: coverletter.tex
\begin{center}
\section*{\large Foreword to the ATLAS and CMS contributions to the DPF Snowmass Process}

\em{\large The ATLAS and CMS Collaborations}
\end{center}

{\large
\noindent In 2012 the CMS and ATLAS experiments at CERN discovered a Higgs boson with a mass of 125-126 GeV. 
This opened a new chapter in the history of particle physics. 
In this brief cover note, the main priorities of the ATLAS and CMS Collaborations in this new era are presented. 
Further adjustments to these priorities may occur as detailed studies of this particle and searches for new physics are extended into new realms at higher energies in the future.

\medskip

\noindent The Higgs discovery was anchored by the final states that afforded the best mass resolution, namely $H\rightarrow \gamma \gamma$ and $H\rightarrow ZZ$ ($4e$, $4\mu$ or $2e2\mu$). 
These modes placed stringent requirements on detector design and performance.
Indeed, the ability to search for the SM Higgs boson over the fully allowed mass range played a crucial role in the conceptual design and benchmarking of the experiments and also resulted in excellent sensitivity to a wide array of signals of new physics at the TeV energy scale. 
Remarkably, the recent discovery came at half the LHC design energy, much more severe pileup, and one third of the integrated luminosity that was originally judged necessary. 
This demonstrates the great value of a bold early conceptual design, a systematic program of development and construction, and a detailed understanding of detector performance, in confronting challenging physics goals.

\medskip

\noindent With data taken in coming years at or near to the design energy of 14 TeV, a broader picture of physics at the TeV scale will emerge with implications for the future of the energy frontier program.
Amongst the essential inputs will be precision measurements of the properties of the Higgs boson and direct searches for new physics that will make significant inroads into new territory. 
For the foreseeable future the LHC together with CMS and ATLAS will be the only facility able to carry out these studies. 
This program is very challenging for the experiments because it requires accurate reconstruction and identification of physics objects (leptons including the $\tau$, heavy flavor tagging, photons, jets, missing transverse energy) from relatively low to very high transverse momenta extending to large rapidity (e.g. to characterize events from vector boson fusion).
To retain and extend these capabilities to higher luminosities in the 2020s, existing systems need to be upgraded or replaced. 
This will require a vision as ambitious as that of the original LHC program, in particular ensuring that sufficient resources, both financial and human, are made available in a timely fashion; for R\&D in the short term, for prototyping in close liaison with industry in the medium term, and further down the line for construction.

\medskip

\noindent To realize the full physics potential of the LHC, it is essential that the High Luminosity (HL-LHC) upgrade to the accelerator be carried out. 
However, the planned instantaneous luminosity of $\sim 5 \times 10^{34}\,\text{cm}^{-2}\text{s}^{-1}$ is well beyond its original design and thus the capabilities of the current experiments. 
This necessitates upgrades/replacements explicitly targeted towards the broad program of physics mentioned above.

\medskip

\noindent The European Strategy for Particle Physics, formally adopted by the CERN Council in May 2013, states that the ``top priority should be the exploitation of the full potential of the LHC, including the high-luminosity upgrade of the machine and detectors with a view to collecting ten times more data than in the initial design, by around 2030.''
The European Strategy clearly recognizes that ``the scale of the facilities required by particle physics is resulting in the globalisation of the field.''

\medskip

\noindent ATLAS and CMS, as worldwide collaborations, are fully committed to engage all international partners to deliver this program. 
The U.S. has made contributions critical to the success of the CMS and ATLAS experiments to date. 
The success of future data-taking, and the detector upgrade programs, will rely on a continuing strong engagement of U.S. groups in the two experiments to assure the continued success of the high energy frontier program.

\medskip

\noindent In summary:

The highest priority in particle physics should be to exploit fully the physics potential of the LHC.
To achieve this goal, ATLAS and CMS place the highest priority on securing the resources needed to achieve the following goals:
\begin{itemize}
\item Upgrade/replace selected elements of the apparatus and associated readout, trigger, data acquisition and computing systems in order to optimally exploit fully the phase of LHC running above the original design luminosity over the next 10 years (to LHC long-shutdown 3, ``LS3'');
\item Prepare, prototype and construct the necessary upgrades/replacements of detectors to operate and optimally exploit the phase of running at instantaneous luminosities in excess of $5\times 10^{34}\text{cm}^{-2} \text{s}^{-1}$ in the roughly ten year period following LS3.
\end{itemize}
CMS and ATLAS strongly recommend that resources be allocated for the HL-LHC to enable the LHC to operate at luminosities significantly higher than the original design.

%% file: introduction.tex
\section{Introduction}
 From 2015 to 2017, the ATLAS experiment will collect about 100 fb$^{-1}$ of data, with a peak instantaneous luminosity of $10^{34}\text{cm}^{-2}\text{s}^{-1}$, and at a center-of-mass energy between 13 and 14 TeV, nearly twice the energy of the 2012 LHC running.
Following this so-called ``Run 2,'' the accelerator will be upgraded to deliver two to three times the instantaneous luminosity at $\sqrt{s}$=14 TeV and the ATLAS detector will undergo the ``Phase-I'' upgrade to maintain the detector capabilities at the increased luminosities.
A total of 300-350 fb$^{-1}$ of data is expected to be collected by the end of Run 3 in 2021. The final focus magnets in the interaction regions will begin to suffer radiation damage by this stage, and a Phase-II upgrade to the LHC is proposed to provide an instantaneous luminosity of $5 \times 10^{34} \text{cm}^{-2}\text{s}^{-1}$ beginning in 2023. The ATLAS experiment will also require upgrades for this High Luminosity LHC (HL-LHC) program, to maintain its capabilities at the planned instantaneous luminosity, which corresponds to an a average of 140 interactions per crossing.
The ATLAS Phase-II upgrades for HL-LHC are described in the Letter of Intent~\cite{PhIILoI} and are summarized below.
A specific subset of instrumentation studies being performed in the U.S. is given in a separate contribution to the Snowmass Community Planning Study~\cite{Brooijmans:2013gba}.

 Hadron colliders have a major role as exploration and discovery machines, and the HL-LHC is required to fully exploit the opportunities at the LHC for discovery of physics beyond the Standard Model (``BSM physics''). Precision studies of the newly discovered Higgs boson, which may provide a portal to BSM physics, and direct searches for additional Higgs particles, are a very high priority at the LHC and the HL-LHC provides a significant increase in discovery reach for new Higgs particles and sensitivity to non-Standard Model effects in Higgs boson couplings. These opportunities also exist in other direct searches for BSM physics, such as supersymmetry (SUSY) or new heavy gauge bosons ($Z^\prime$ and $W^\prime$).

In this Whitepaper we summarize studies that examine the physics capabilities of ATLAS at the HL-LHC for a number of key physics processes.
These studies were started in the context of developing the European Strategy for Particle Physics~\cite{ATL-PHYS-PUB-2012-001, ATL-PHYS-PUB-2012-004}, a process that affirmed the HL-LHC program as a top priority in particle physics.
This note summarizes both the results used as input for the European Strategy and results that have been updated since then.
We contrast the projected results at two benchmark values of integrated luminosity: the 300 fb$^{-1}$ expected by the end of Run 3, and the 3000 fb$^{-1}$ expected to be delivered by the HL-LHC.
These projections are based on extrapolating the current detector performance from the current pile-up conditions in the data of an average number, $\mu$, of 20 interactions per crossing, and from simulation of pile-up for Run 3 with up to $\mu=69$, to the conditions at HL-LHC of $\mu=140$. The parameterizations used for the extrapolation are described in detail in Ref.~\cite{PUB-2013-004}. These parameterizations provide rather conservative estimates of the reach and precision of measurements. Except where otherwise noted, they do not include improvements due to new techniques, improved understanding of backgrounds, or reduced theoretical uncertainties.

%% file: detectors.tex
\section{Detector Requirements and Upgrade Designs}

The promise of the rich physics program at the HL-LHC cannot be fulfilled without essential improvements to the ATLAS apparatus.
In particular, the studies summarized below depend on robust tracking peformance, moderate trigger rates for single leptons, good resolution on missing transverse energy measurements, and enhanced jet tagging and $\tau$ identification.
The ATLAS Phase-II upgrades are intended to maintain and, in some cases, improve the detector performance during high-luminosity operation.

The current ATLAS detector was designed to operate efficiently for an integrated luminosity of $300\,\ifb$ and a pileup rate of 20-25 events per bunch crossing at the design luminosity of $1 \times 10^{34} \, \text{cm}^{-2}\text{s}^{-1}$.
The HL-LHC will feature integrated luminosities and pileup rates well beyond these values.

The ATLAS Inner Detector will need to be replaced as the accumulated radiation dose starts to damage the silicon detectors, and the occupancy in some regions of the silicon and in the straw tubes will be unacceptably high.
An all-silicon inner tracker is proposed, using high-granularity radiation-hard sensors that can withstand the particle fluences expected during the high-luminosity run.
The new tracker will provide a sufficient number of space points for track reconstruction for ATLAS to maintain high tracking efficiency and a low fake rate in the increased pileup environment.
The ATLAS calorimeters will be upgraded with new front-end electronics that make it possible to digitize the response in every cell, so that the digitized data can be used in fast trigger algorithms.
In a similar way, the muon spectrometer will be upgraded with new front-end electronics to read out the chambers in time for a fast trigger decision.

The new trigger architecture is intended to operate with a 500-kHz Level-0 decision that uses the information from the fast readouts of the muon spectrometer and calorimeters before adding tracking information at Level-1.
ATLAS submitted its Letter of Intent for the Phase-II Upgrade~\cite{PhIILoI} in March 2013, and detailed Technical Design Reports are being prepared for the various detector systems to be upgraded.

\section{Parameterization of Expected Performance}
\label{sec:detectorperformance}
The studies of the ATLAS high-luminosity physics program are based on a set of performance assumptions for the reconstructed physics objects~\cite{PUB-2013-004}.
These parameterizations leverage the excellent understanding of the current ATLAS detector and recent full simulations of the upgraded systems.
For the most part, they are based on the Run 1 detector with conservative realistic assumptions on the pileup dependence.
The missing transverse momentum resolution is extrapolated from studies with 7 and 8 TeV data and full simulation.
It is parameterized as a function of the number of pileup events, which naturally increase at high instantaneous luminosity.
These results are quite conservative for the most part.
For example, a comparison of the b-tagging performance assumption and the results from full simulation is shown in Fig.~\ref{fig:btagging}, in which the upgraded performance is shown to be better than the performance inferred from the current b-tagging with high pileup, due to the inner tracker improvements.
In particular, a b-tagging algorithm that is 75\% efficient was assumed to to have a light-jet rejection factor of 30 for the physics studies presented here, when in fact a full simulation study points to a light-jet rejection factor of approximately 120.

\begin{figure}[htbp]
\begin{center}
\includegraphics[width=0.48\textwidth]{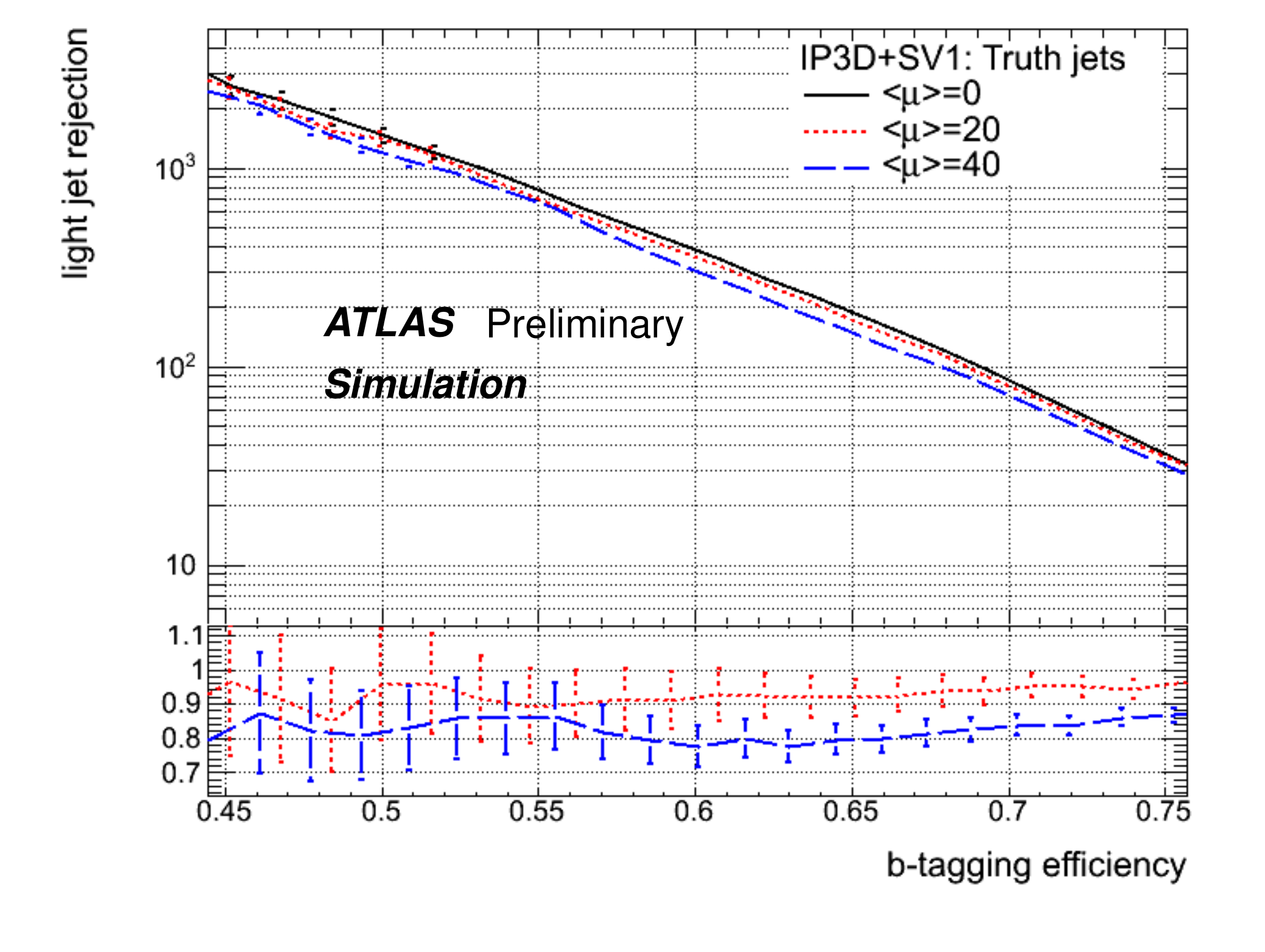}
\includegraphics[width=0.48\textwidth]{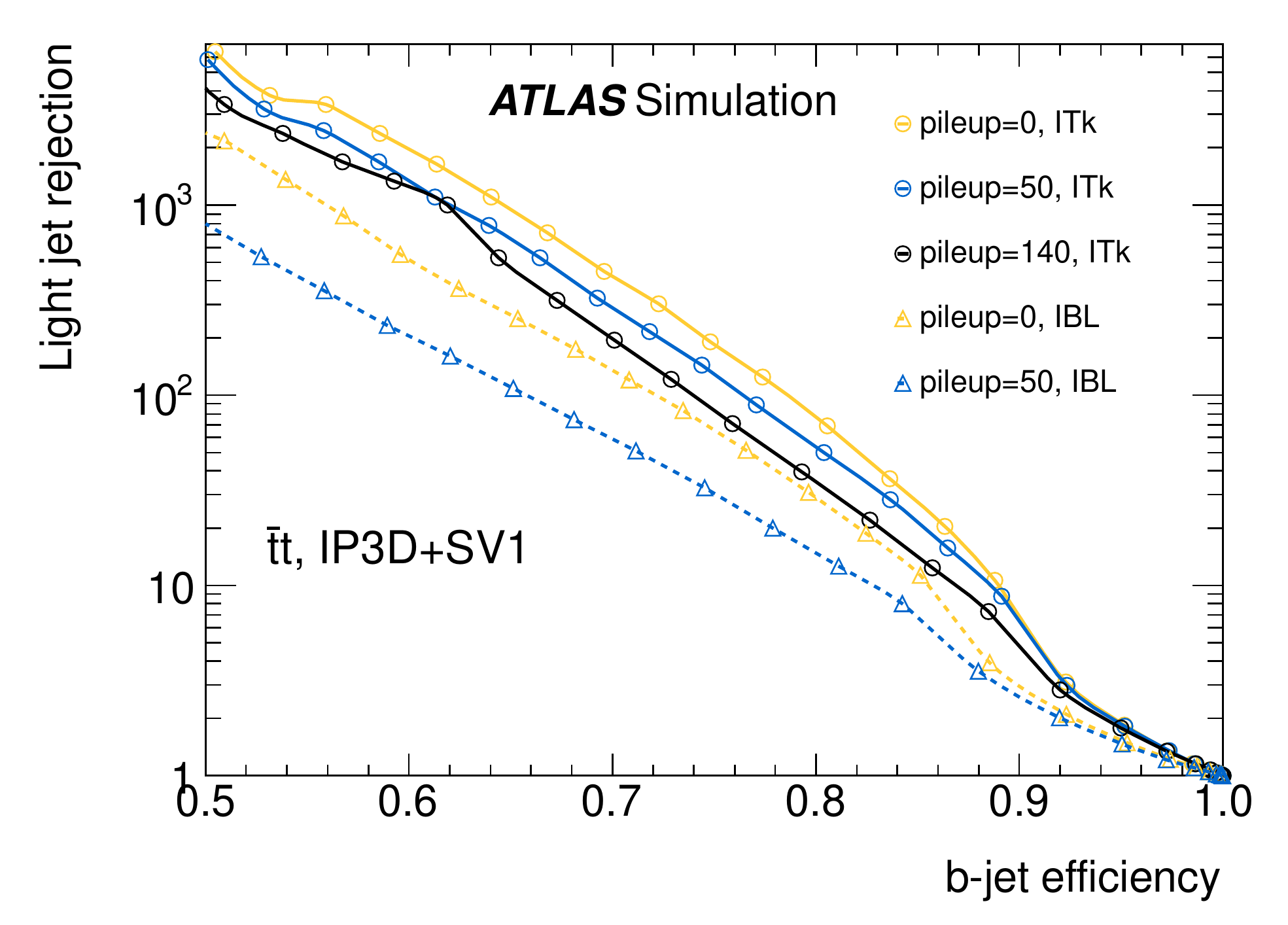}
\end{center}
\caption{Comparison of the ATLAS b-tagging performance parameterization as a function of pileup (left) and the b-tagging performance in full simulation for various pileup scenarios and detector configurations (right).  The IP3D+SV1 tagging algorithm uses a combination of 3-dimensional impact parameter likelihood and secondary vertexing to achieve high performance, especially when the Insertable B-Layer pixel detector (IBL) or proposed all-silicon Inner Tracker (ITK) are used.}
\label{fig:btagging}
\end{figure}

%% file: higgs.tex
\section{Measurements of the Higgs boson}
\label{sec:Higgs} With the discovery of a Higgs
boson~\cite{Aad:2012tfa,Chatrchyan:2012ufa} last summer, a major
focus of the LHC program has become the measurement of the properties
of this new particle.  The resonance was termed a ``Higgs-like'' boson
when only its mass was known with any precision.  After another year
of study, $J^P=0^+$ is strongly favored~\cite{Chatrchyan:2012jja, 2013HiggsJP}.  With
limited precision, the new particle's couplings agree with SM
expectations in those channels for which they have been
measured~\cite{2013HiggsCoupling}.  The ratio of the combined signal strength to the Standard Model value is now measured by ATLAS to be $1.33\pm0.21$.
There is no significant indication of a deviation from the
properties of a SM Higgs boson, but the precision of the coupling
measurements leaves room for BSM physics, for which models typically
predict deviations from the SM couplings.  However, the deviations
can be arbitrarily small~\cite{PhysRevD.86.095001}, as, for example, in SUSY scenarios where the
additional Higgs particles are very heavy.  Thus it is of the utmost
importance to measure the couplings with increased precision, and to
search directly for additional Higgs particles.

\subsection{Higgs boson couplings}
\label{sec:HiggsCouplings} Because the mass of the observed Higgs boson is
approximately 125 GeV, a large variety of decay channels are open to
investigation at the LHC.  The LHC experiments measure the product
of the production cross section and the branching fraction into a
particular final state.  In order to extract the Higgs boson
couplings and test the Standard Model predictions, fits to the
measured signal strengths are done using all relevant
channels~\cite{2013HiggsCoupling}. Achieving the best sensitivity to
potential non-Standard Model Higgs boson couplings requires precision
measurements of as many different Higgs production and decay
channels as possible.

Current projections to the HL-LHC are based primarily on those
channels that are measured in the current dataset, with the addition
of a few key rare channels that can be accessed only at the HL-LHC~\cite{ATL-PHYS-PUB-2012-004}.
The current analyses measure Higgs production through both the
gluon-fusion and vector boson fusion (VBF) channels into final
states $\gamma\gamma$, $ZZ^*$, and $WW^*$, and good progress is being made towards measurements in $\tau^+\tau^-$ and $b\bar{b}$. The
luminosity of the HL-LHC will provide improved statistical precision
for already established channels and allow rare Higgs boson
production and decay modes to be studied and measured with
substantially improved precision compared to the measurements that
will be made by ATLAS with about 300 fb$^{-1}$ of data (Run 3).

Changes to the trigger and to the photon and lepton selections that
are needed at high-luminosity to keep rates in check are taken into
account. For the VBF jet selection, the cuts were tightened to
reduce the expected fake rate induced by pile-up to below 1\% of the
jet activity from background processes.

The following channels, which are already studied in the current 7
and 8 TeV datasets, are now evaluated for the 14 TeV HL-LHC dataset:

\begin{itemize}
\item $H \to \gamma\gamma$ in the 0-, 1-, and 2-jet final states.
The analysis is carried out analogously to Ref.~\cite{Aad:2012tfa}.
\item Inclusive $H \to ZZ^{*} \to 4\ell$, following a selection close to that in Ref.~\cite{Aad:2012tfa}.
\item $H \to WW^{*}\to \ell\nu\,\ell\nu$ in the 0-jet and the 2-jet final state, the latter with a VBF selection.  The analysis follows closely that of Ref.~\cite{Aad:2012tfa}.
\item $H \to \tautau$ in the 2-jet final state with a VBF selection as in Ref.~\cite{2011HtautauPaper}.
\end{itemize}

The $WW^*$ and $\tau^+\tau^-$ channels are challenging as they require a detailed understanding of the various backgrounds.  The ultimate precision will depend on how well these backgrounds can be constrained, {\em in situ}, using data.  For this study it was assumed that the background understanding will not improve beyond what was achieved by summer 2012.  The projections are therefore rather pessimistic.  For example, for the $WW^*$ channel the background uncertainty is already significantly improved using the full 2012 dataset, primarily due to improved analysis techniques~\cite{2013HiggsCoupling}, resulting in a precision on $\mu$ of $\approx 30$\%.  These improvements have not yet been propagated into the current study, and therefore it should be possible to improve on the quoted precision of 29\%.

To exploit the projected 3000 fb$^{-1}$ provided by the
HL-LHC, several additional, relatively rare, channels with Higgs boson decays
into the high-resolution final states $H \to \gamma\gamma$ and $H \to \mu\mu$ are studied:

\begin{itemize}
\item $t\bar{t}H, H \to \gamma\gamma$ and $H \to \mu\mu$.
\item $WH/ZH, H \to \gamma\gamma$.
\item Inclusive $H \to \mu\mu$.
\end{itemize}

\begin{figure}[!h]
\begin{center}
  \includegraphics[width=0.48\linewidth]{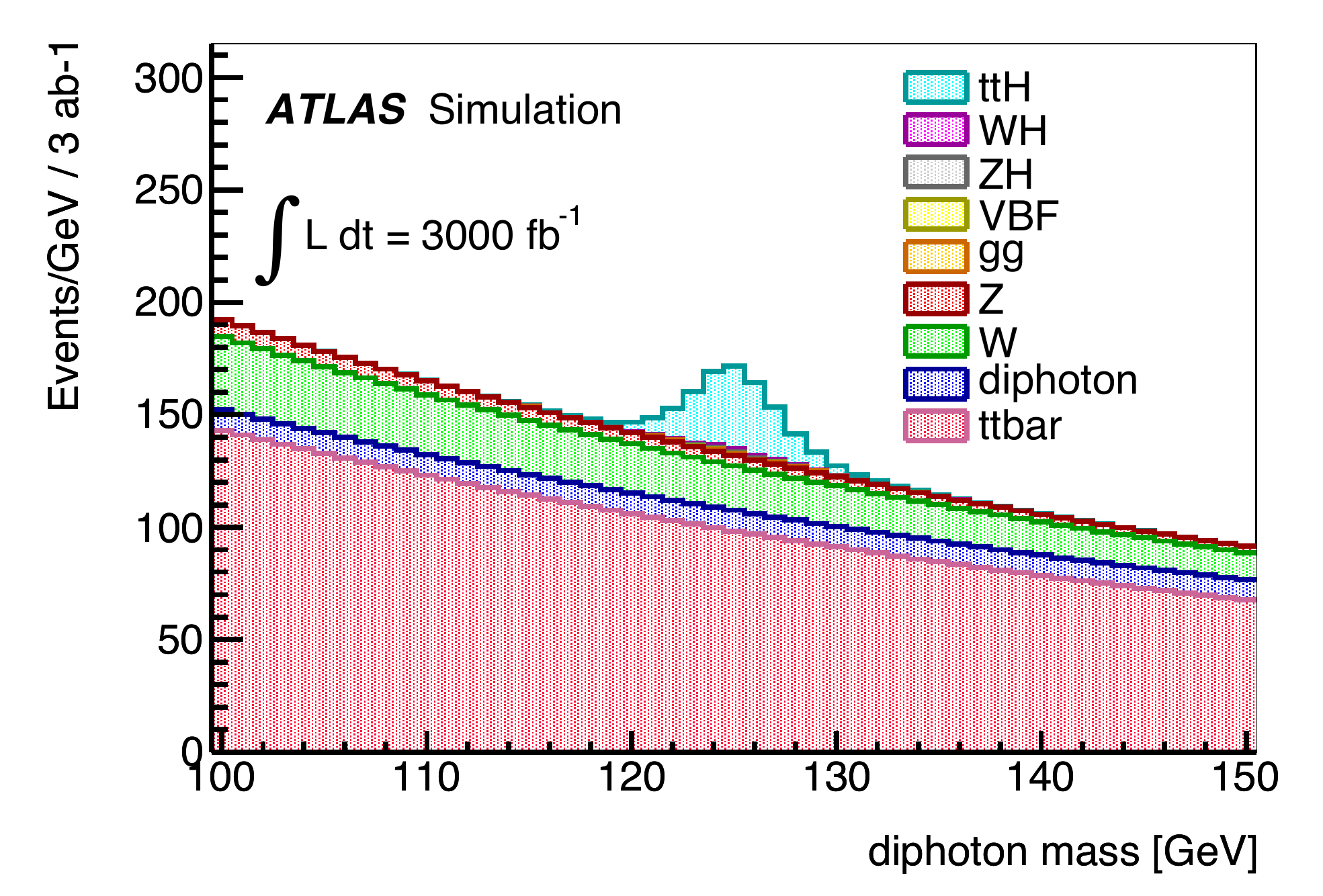}
  \caption{Expected diphoton mass distribution in the single lepton ttH channel for $\sqrt{s}$=14 TeV and $\mathcal{L}=3000\,\text{fb}^{-1}$.}
\label{fig:tth_gamgam}
\end{center}
\end{figure}

\begin{figure}[!h]
\begin{center}
  \includegraphics[width=0.48\linewidth]{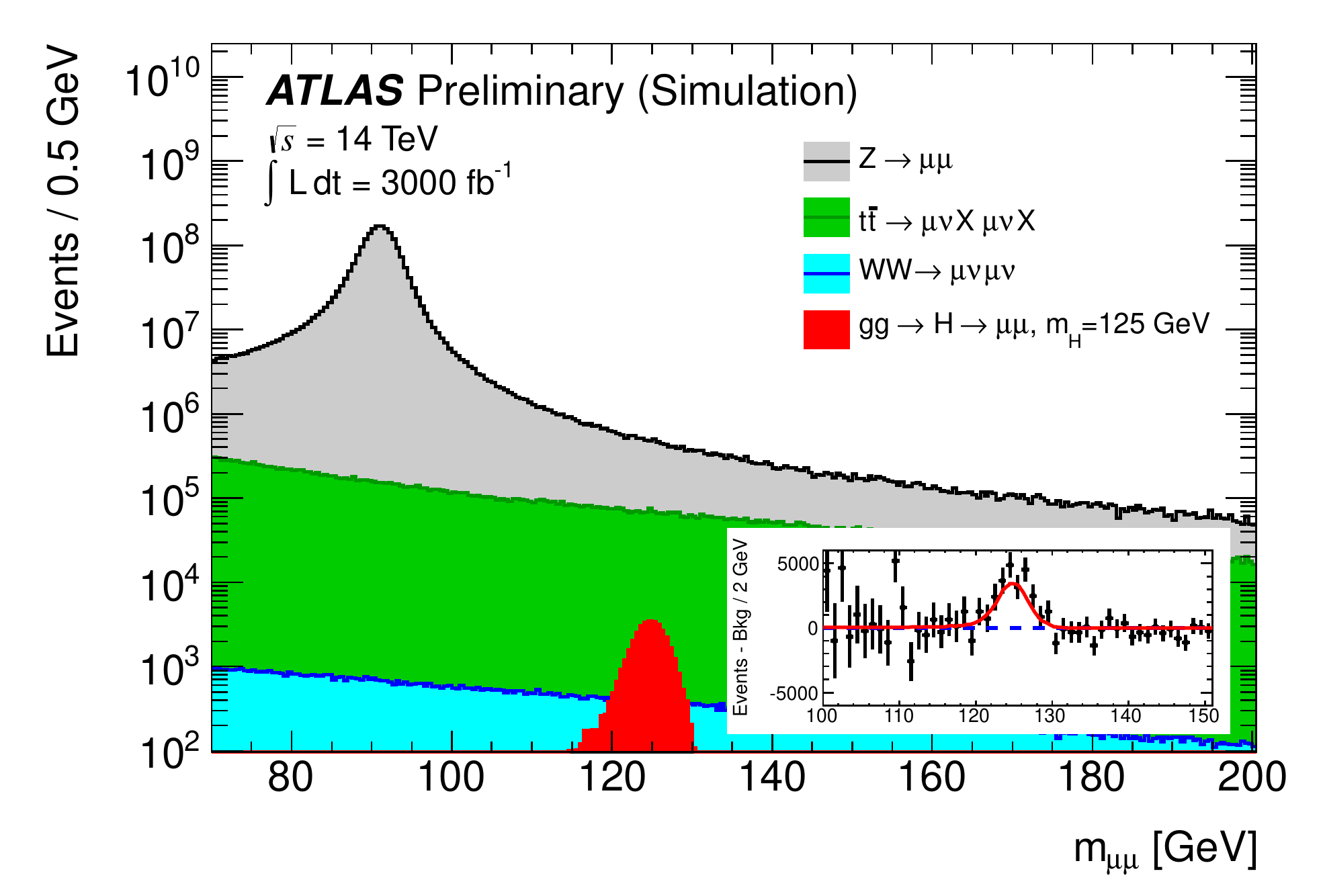}
  \caption{Distribution of the $\mu^+\mu^-$ invariant mass of the signal and background
  processes generated for $\sqrt{s}$=14 TeV and $\mathcal{L}=3000\,\text{fb}^{-1}$.}
\label{fig:h_mumu}
\end{center}
\end{figure}

The $t\bar{t}H$ and $WH/ZH$ $\gamma\gamma$ channels above have a low
signal rate at the LHC, but one can expect to observe more than 100
signal events with the HL-LHC.  The selection of the diphoton system
is done in the same way as for the inclusive $H \to \gamma\gamma$
channel. In addition, 1- and 2-lepton selections, dilepton mass cuts
and different jet requirements are used to separate the $WH$, $ZH$
and $t\overline{t}H$ initial states from each other and from the
background processes. The $t\overline{t}H$ initial state gives the
cleanest signal with a signal-to-background ratio of $\sim$20\%, to
be compared to $\sim$10\% for $ZH$ and $\sim$2\% for $WH$.  The $H
\to \mu\mu$ decay will be measured for the first time in the Run 3,
300 fb$^{-1}$ dataset, but only with very limited precision.  The
HL-LHC will allow a measurement in the inclusive channel of better
than 20\% and in $t\bar{t}H, H \to \mu\mu$ of just over 20\%.
The expected $H
\to \gamma\gamma$ signal in $t\overline{t}H$ for 3000 fb$^{-1}$ is
shown in Fig.~\ref{fig:tth_gamgam}, and the inclusive $H \to \mu\mu$
expectation is shown in Fig.~\ref{fig:h_mumu}.

It is interesting to measure the ratio of
third-generation to second-generation couplings, to test the Standard Model Higgs boson couplings and the potential for BSM effects.
Studies focus on VBF production with  decays to $\tau\tau\rightarrow\ell\tau_{had} 3\nu$ and
$\tau\tau\rightarrow\ell\ell^\prime 4\nu$.
A relatively precise $H \to \mu\mu$ signal-strength
measurement and improved $H \to \tau\tau$ measurements provide a
significant improvement in the measurement of the ratio of partial
widths into second- and third-generation fermions.

\begin{figure}[!h]
\begin{center}
  \begin{minipage}{0.45\linewidth}
 {\includegraphics[width=0.98\linewidth]{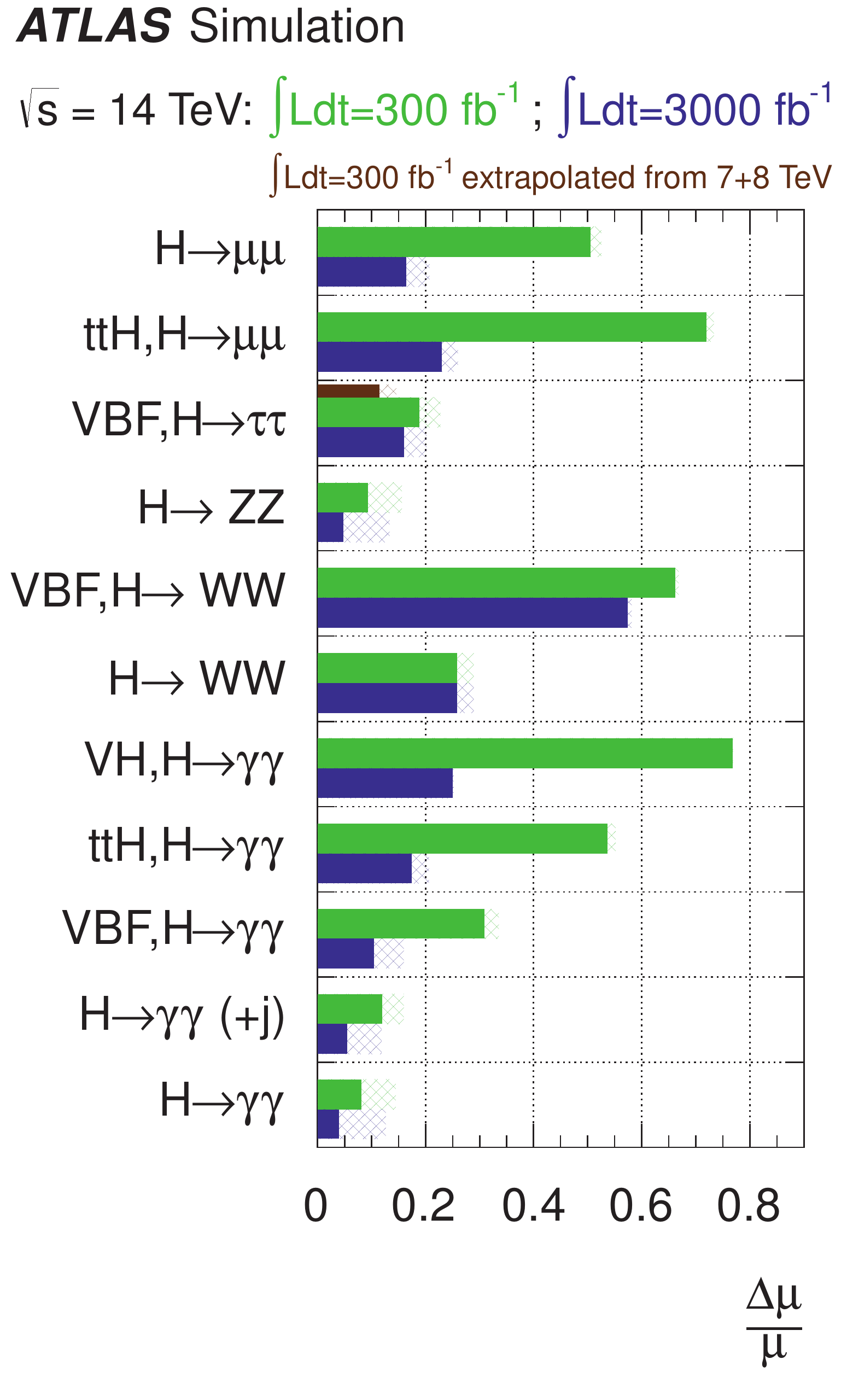} }%
\begin{center} (a) \end{center}
  \end{minipage}
  \hspace{1cm}
  \begin{minipage}{0.45\linewidth}
{\includegraphics[width=0.98\linewidth]{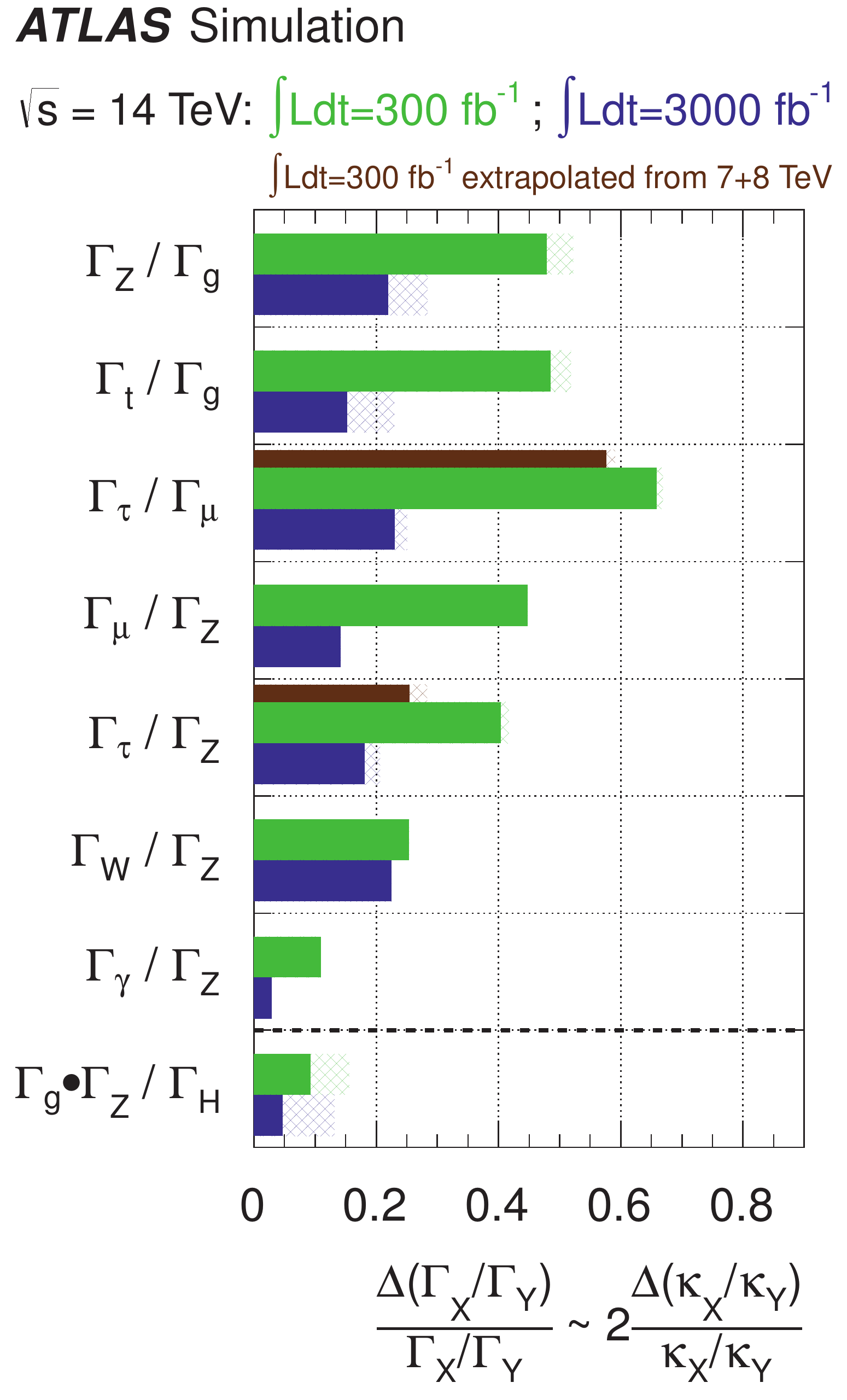} }%
\begin{center} (b) \end{center}
  \end{minipage}
\end{center}
\caption{Summary of Higgs analysis sensitivities wth 300~\ifb and 3000~\ifb at $\sqrt{s}=14$ TeV for a SM Higgs boson with a mass of
  125~\GeV. Left: Uncertainty on the signal strength.  For the $H \to \tau\tau$ channels the thin brown bars show the expected precision reached from extrapolating all tau-tau channels studied in the current 7 TeV and 8 TeV analysis to 300 fb$^{-1}$, instead of using the dedicated studies at 300 fb$^{-1}$ and 3000 fb$^{-1}$ that are based only in the VBF $H \to \tau\tau$ channels. Right: Uncertainty on ratios of partial decay width fitted to all
  channels. The hashed areas indicate the increase of the estimated error due
  to current theory systematic uncertainties.}
\label{fig:coupling_fits}
\end{figure}

The projected uncertainties on the signal-strength measurements and
the ratios of partial widths are summarized in
Fig.~\ref{fig:coupling_fits} for the channels that have been studied
to date.
The same uncertainties are given in tabular form in Tables~\ref{tab:higgs_mu_300} and \ref{tab:higgs_mu_3000}.
The partial-width ratio uncertainties are given as well in
Table~\ref{tab:gammafit}.

\begin{table}[h]
\centering
\begin{tabular}{l|l|l}
\hline
 & with theory systematics & without theory systematics \\
\hline
  $H\rightarrow\mu\mu$ & 0.53 & 0.51 \\
  $ttH,H\rightarrow\mu\mu$ & 0.73 & 0.72 \\
  $VBF,H\rightarrow\tau\tau$ & 0.23 & 0.19 \\
  $VBF,H\rightarrow\tau\tau$ (extrap) & 0.15 & 0.11 \\
\hline
  $H\rightarrow ZZ$ & 0.16 & 0.093 \\
  $VBF,H\rightarrow WW$ & 0.67 & 0.66 \\
  $H\rightarrow WW$ & 0.29 & 0.26 \\
\hline
  $VH,H\rightarrow\gamma\gamma$ & 0.77 & 0.77 \\
  $ttH,H\rightarrow\gamma\gamma$ & 0.55 & 0.54 \\
\hline
  $VBF,H\rightarrow\gamma\gamma$ & 0.34 & 0.31 \\
  $H\rightarrow\gamma\gamma (+j)$ & 0.16 & 0.12 \\
  $H\rightarrow\gamma\gamma$ & 0.15 & 0.081 \\
\hline
\end{tabular}
\caption{Expected relative uncertainties on the signal strength $\mu$ for $300$ fb$^{-1}$. The $H\rightarrow\tau\tau$ line labeled `(extrap)' is based on an extrapolation to $300$ fb$^{-1}$ from all $\tau\tau$ channels currently studied in the 7 TeV and 8 TeV analyses, whereas the other $\tau\tau$ projection is based on dedicated studies based only on the VBF production channel. }
\label{tab:higgs_mu_300}
\end{table}

\begin{table}[h]
\centering
\begin{tabular}{l|l|l}
\hline
 & with theory systematics & without theory systematics \\
\hline
  $H\rightarrow\mu\mu$ & 0.21 & 0.16 \\
  $ttH,H\rightarrow\mu\mu$ & 0.26 & 0.23 \\
  $VBF,H\rightarrow\tau\tau$ & 0.20 & 0.16 \\
\hline
  $H\rightarrow ZZ$ & 0.13 & 0.047 \\
  $VBF,H\rightarrow WW$ & 0.58 & 0.57 \\
  $H\rightarrow WW$ & 0.29 & 0.26 \\
\hline
  $VH,H\rightarrow\gamma\gamma$ & 0.25 & 0.25 \\
  $ttH,H\rightarrow\gamma\gamma$ & 0.21 & 0.17 \\
\hline
  $VBF,H\rightarrow\gamma\gamma$ & 0.16 & 0.11 \\
  $H\rightarrow\gamma\gamma (+j)$ & 0.12 & 0.054 \\
  $H\rightarrow\gamma\gamma$ & 0.13 & 0.040 \\
\hline
\end{tabular}
\caption{Expected relative uncertainties on the signal strength $\mu$ for $3000$ fb$^{-1}$.}
\label{tab:higgs_mu_3000}
\end{table}

\begin{table}[htbp]
  \centering
  \begin{tabular}{|l|l|l|l|l|}
\hline
 & \multicolumn{2}{|c|}{ 300\,\ifb}& \multicolumn{2}{|c|}{ 3000\,\ifb} \\
\hline
     & w/theory uncert. & wo/theory uncert. & w/theory uncert. & wo/theory uncert.\\
\hline
  $\Gamma_{Z} / \Gamma_{g}$                  & 0.52 & 0.48 & 0.28 & 0.22 \\
  $\Gamma_{t} / \Gamma_{g}$                  & 0.52 & 0.49 & 0.23 & 0.15 \\
\hline
  $\Gamma_{\tau} / \Gamma_{\mu}$              & 0.67 & 0.66 & 0.25 & 0.23  \\
  $\Gamma_{\tau} / \Gamma_{\mu}$ (extrap)     & 0.59 & 0.58 &      &       \\
\hline
  $\Gamma_{\mu} / \Gamma_{Z}$                & 0.45 & 0.45 & 0.14 & 0.14 \\
  $\Gamma_{\tau} / \Gamma_{Z}$               & 0.42 & 0.40 & 0.21 & 0.18 \\
  $\Gamma_{\tau} / \Gamma_{Z}$ (extrap)      & 0.28 & 0.26 &      &       \\
  $\Gamma_{W} / \Gamma_{Z}$                  & 0.25 & 0.25 & 0.23 & 0.23 \\
  $\Gamma_{\gamma} / \Gamma_{Z}$              & 0.11 & 0.11  & 0.029 & 0.029 \\
\hline
  $\Gamma_{g}\bullet\Gamma_{Z} / \Gamma_{H}$ & 0.16 & 0.093  & 0.13 & 0.047 \\
\hline
  \end{tabular}
  \caption{Relative uncertainty on the ratio of partial widths
    for the combination
    of Higgs analysis and coupling properties fits at
    14~\TeV, 300~\ifb\ and 3000~\ifb, assuming a SM Higgs Boson with a mass
    of 125~\GeV.}
  \label{tab:gammafit}
\end{table}

\begin{figure}[h!]
\begin{center}
\includegraphics[width=0.48\textwidth]{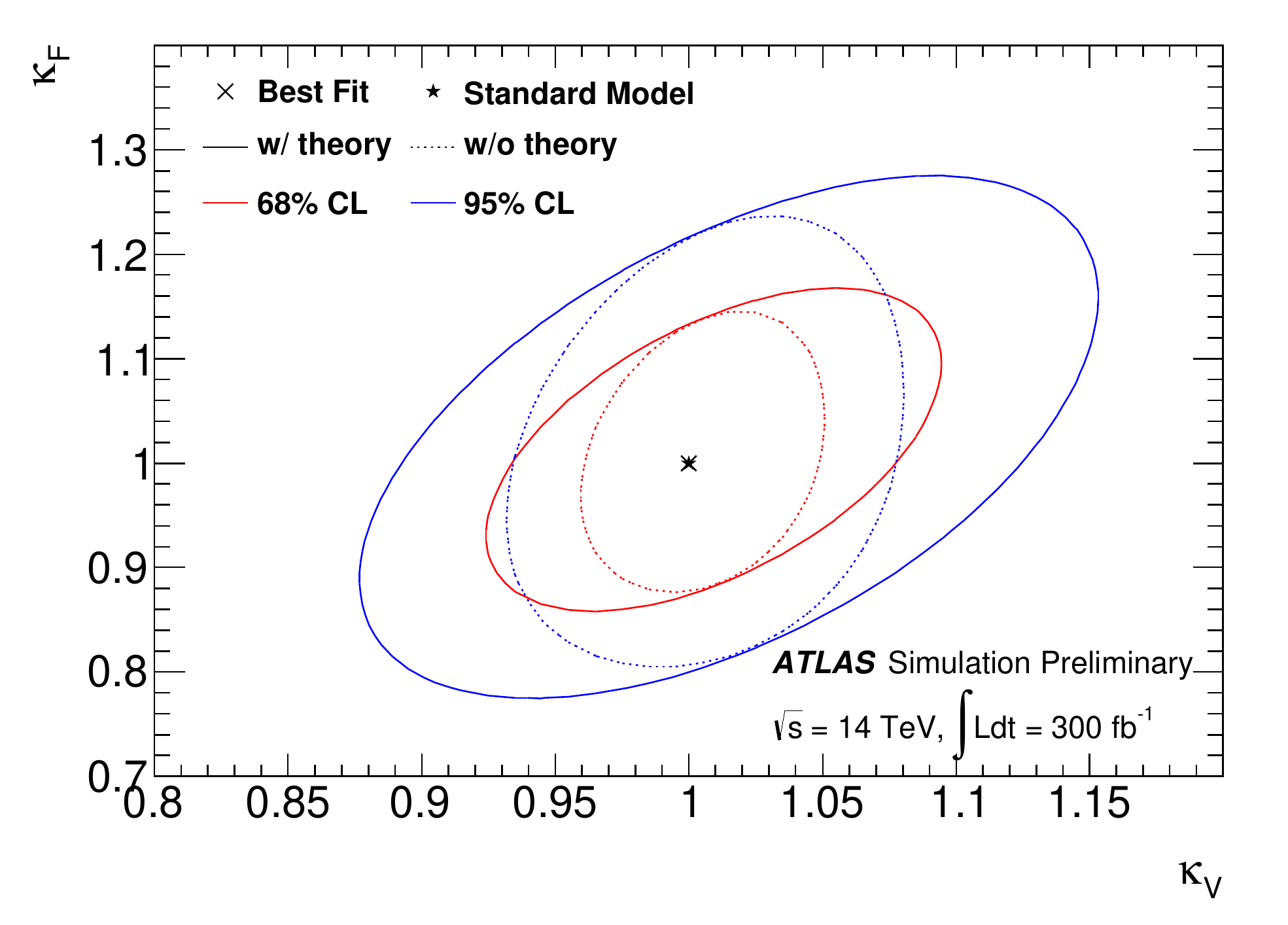}
\includegraphics[width=0.48\textwidth]{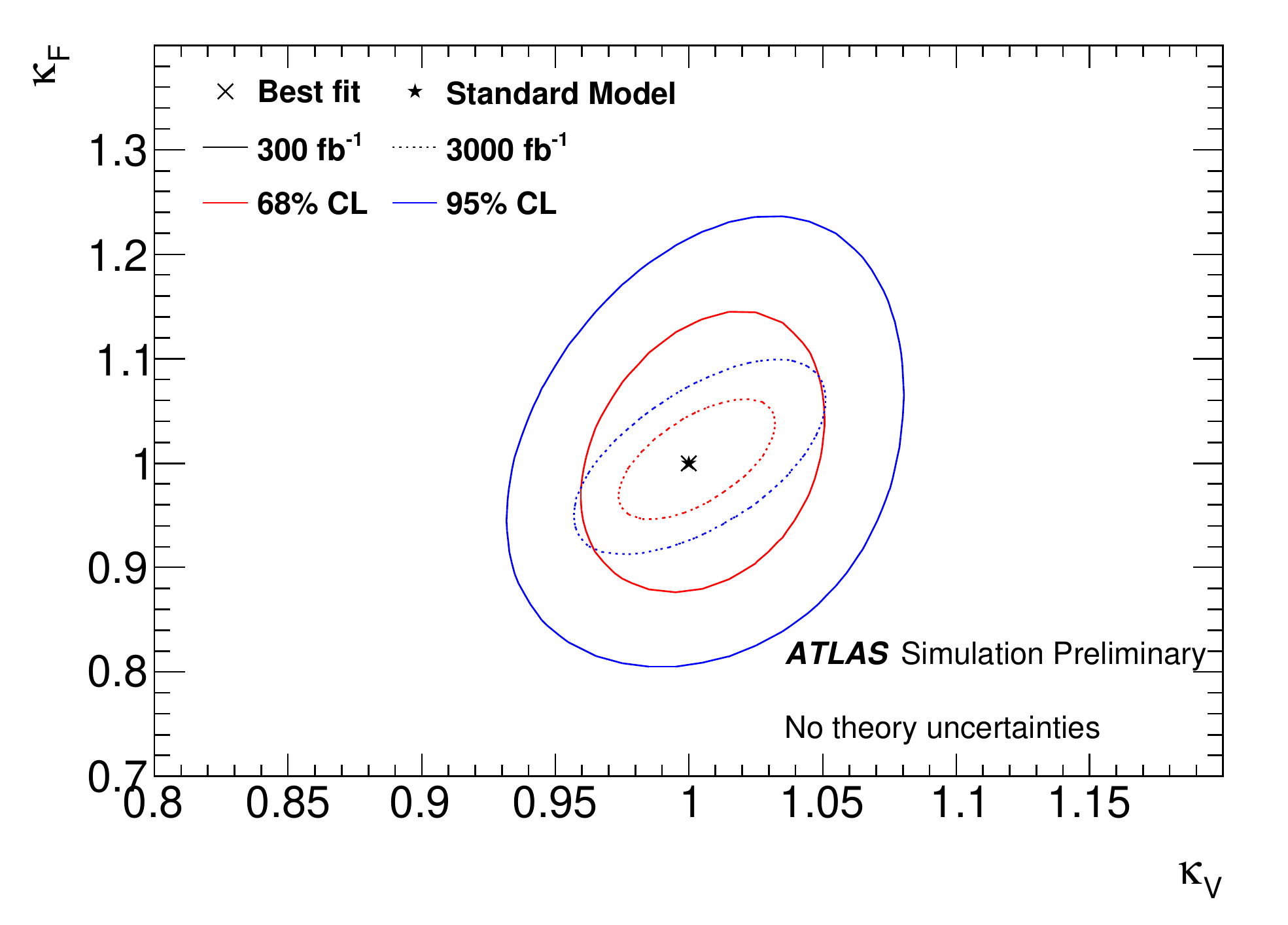}
\end{center}
\caption{68\% and 95\% confidence level (CL) likelihood contours for $\kappa_V$ and $\kappa_F$ in a minimal coupling fit at 14~\TeV.
Left: impact of the theory uncertainties for an assumed integrated luminosity of 300~\ifb.
Right: results without theory uncertainties for 300~\ifb\ and 3000~\ifb.
\label{fig:kVkF}
}
\end{figure}

The ratios of partial widths shown in the right-hand panel of
Fig.~\ref{fig:coupling_fits} correspond to coupling scale-factors
according to $\Gamma_X/\Gamma_Y = \kappa_X^2/\kappa_Y^2$, where
$\kappa_i$ is the coupling scale-factor for the Higgs
coupling\footnote{In the case of gluons and photons, these are
effective couplings that include all loop effects into a single
value} to $i=g,\gamma,W,Z,t,\mu,\tau$, and the Standard Model value
is $\kappa=1$~\cite{HiggsXsecWG, Heinemeyer:2013tqa}.
The results of a minimal fit,
where only two independent scale factors are used, $\kappa_V$ for
vector bosons and $\kappa_F$ for fermions, is shown in
Table~\ref{tab:couplings_CVCF}.  Significant improvement in the
precision between 300 and 3000 fb$^{-1}$ is seen. The
column including the theory uncertainty assumes no improvement over
today's values, certainly a pessimistic assessment.
Fig.~\ref{fig:kVkF} shows the two-dimensional contours in $\kappa_V$
and $\kappa_F$.  The left-hand figure compares the projected results
for 300 fb$^{-1}$ with, and without, theory uncertainties included.
The right-hand figure compares the 300 fb$^{-1}$ and 3000 fb$^{-1}$
results with no theory uncertainties included.

\begin{table}[h!tb]
\begin{center}
\renewcommand{\arraystretch}{1.2}
\begin{tabular}{|l|c|c|}
\hline
Coupling & With theory systematics & Without theory systematics \\\hline
\multicolumn{3}{|c|}{300~\ifb}\\\hline
$\kappa_V$ & $^{+5.9\%}_{-5.4\%}$  & $^{+3.0\%}_{-3.0\%}$\\ \hline
$\kappa_F$ & $^{+10.6\%}_{-9.9\%}$ & $^{+9.1\%}_{-8.6\%}$ \\ \hline
\multicolumn{3}{|c|}{3000~\ifb}\\\hline
$\kappa_V$ & $^{+4.6\%}_{-4.3\%}$ & $^{+1.9\%}_{-1.9\%}$ \\ \hline
$\kappa_F$ & $^{+6.1\%}_{-5.7\%}$  & $^{+3.6\%}_{-3.6\%}$ \\ \hline
\end{tabular}
\renewcommand{\arraystretch}{1.0}

\caption{Results for $\kappa_V$ and $\kappa_F$ in a minimal coupling fit at 14~\TeV, 300~\ifb\ and 3000~\ifb.}
\label{tab:couplings_CVCF}
\end{center}
\end{table}

\subsubsection{Sensitivity to the Higgs self-coupling}

\begin{figure}[!h]
\begin{center}
  \includegraphics[width=0.75\linewidth]{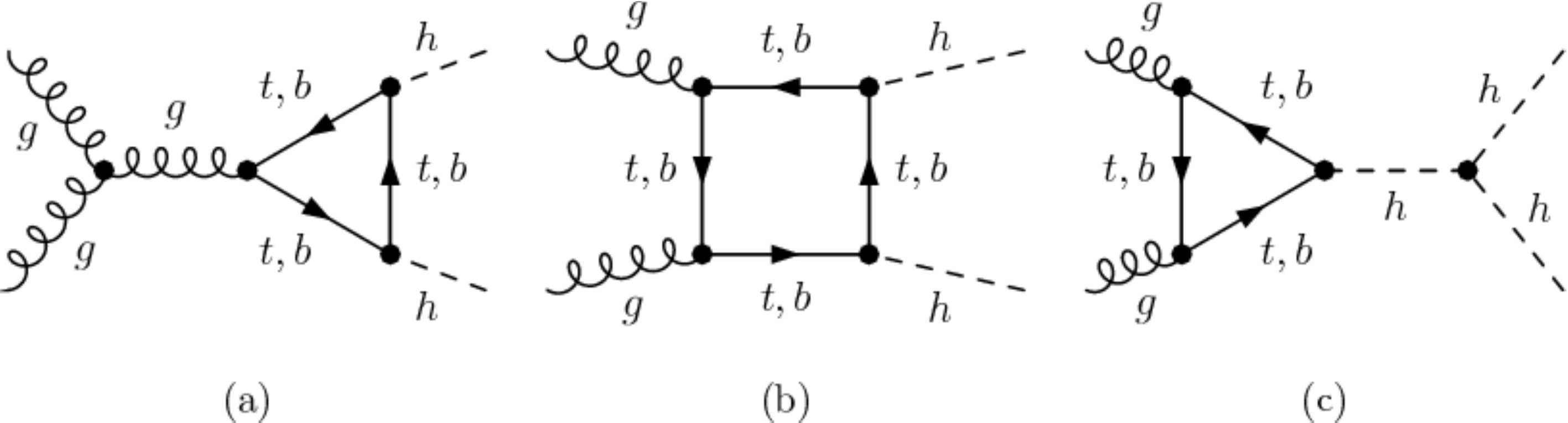}
  \caption{Feynman diagrams for Higgs pair production.}
\label{fig:hh}
\end{center}
\end{figure}

An important feature of the Standard Model Higgs boson is its
self-coupling.  The tri-linear self-coupling $\lambda_{HHH}$ can be
measured through an interference effect in Higgs boson pair
production.  At hadron colliders, the dominant production mechanism
is gluon-gluon fusion.  At $\sqrt{s}=14$ TeV, the production cross
section of a pair of 125 GeV Higgs bosons is estimated at NLO to
be\footnote{The cross section is calculated using the HPAIR
package~\cite{HPAIR}.  Theoretical uncertainties are provided by
Michael Spira in private communication.} $34^{+18\%}_{-15\%}\text{(QCD
scale)}\pm 3\%\text{(PDFs)}\ \text{fb}$.  Figure~\ref{fig:hh} shows the three
contributing diagrams in which the last diagram, the only one that
depends on $\lambda_{HHH}$, interferes destructively with the first
two.  The cross section is therefore enhanced at lower values of
$\lambda_{HHH}$.  For $\lambda_{HHH}/\lambda^{SM}_{HHH}=0~(2)$ the
cross section is 71 (16) fb.  Studies using Higgs pair decays to
$b\overline{b}\gamma\gamma$ and $b\overline{b}W^+W^-$ are in
progress.

%% file: vectorbosons.tex
\section{Measurements of Vector Boson Scattering and Gauge Couplings}

A major reason for expecting new particles or interactions at the TeV energy scale has been the prediction that an untamed rise of the vector boson scattering (VBS) cross section in the longitudinal mode would violate unitarity at this scale.
In the SM it is the Higgs boson which is responsible for the damping of this cross section.
It is important to confirm this effect experimentally, now that one Higgs boson has been observed via direct production and decay.
Alternate models such as Technicolor and little Higgs have been postulated which encompass TeV-scale resonances and a light scalar particle.
These and other mechanisms would modify the vector boson scattering as long as there is a coupling of the new particles to the vector bosons.

The combination of vector boson scattering measurements, triboson production measurements, and Higgs coupling measurements offers a comprehensive program for exploring the gauge-Higgs sector in detail.
For example, measuring vector boson scattering precisely at high mass scales provides sensitivity to new particles and interactions in the electroweak sector.

We summarize results from four studies quantifying the sensitivity to new physics in this sector~\cite{ATL-PHYS-PUB-2013-006}.
The specific studies are $WZ$ VBS in the three-lepton channel, $ZZ$ VBS in the four-lepton channel, $WW$ VBS in the same-sign dilepton channel, and $Z\gamma\gamma$ production in the dilepton plus diphoton channel.

Unlike previous studies that focused on anomalous couplings in a unitarized Higgsless theory~\cite{ATL-PHYS-PUB-2012-005}, these studies are presented in the framework of higher-dimension operators in an effective electroweak field theory~\cite{Degrande:2012wf}.
Multiboson production is modified by certain general dimension-6 and dimension-8 operators containing the Higgs and/or gauge boson fields.  Several representative operators have been chosen to study as benchmarks.  Because higher-dimension operators, as approximations of an underlying $UV$-safe theory, ultimately violate unitarity at sufficiently high energy, care is taken in these studies to select only events in a kinematic range within the unitarity bound, $\Lambda_{UV}$.
These new operators affect only triboson production and vector boson scattering (VBS), but they do not affect other diboson production mechanisms.

The common experimental feature in the following studies of vector boson scattering is the presence of high-$p_T$ jets in the forward-backward regions, similar to those found in Higgs production via vector boson fusion.
The absence of color exchange in the hard scattering process leads to large rapidity intervals with no jets in the central part of the detector; however the rapidity gap topology will be difficult to exploit due to the high level of pileup at a high-luminosity LHC.

\subsection{Vector Boson Scattering}
The selection for VBS studies requires leptons with $p_T > 25\,\text{GeV}$ and, to reduce non-VBS production, at least two high-$p_T$ ($>50\,\text{GeV}$) forward jets are required with an invariant mass of the two highest $p_T$ jets required to be greater than $1\,\text{TeV}$.  In each of the studies below, a particular higher dimension operator is chosen for analysis, but in general each of the VBS channels studied have sensitivity to each of these higher-dimension operators.

The scattering process $ZZ \rightarrow \ell\ell\ell\ell$ is sensitive to the dimension-6 operator 
\begin{equation*}
    {\cal L}_{\phi W} = \frac{c_{\phi W}}{\Lambda^2} {\rm Tr} (W^{\mu \nu} W_{\mu \nu}) \phi^\dagger \phi.
\end{equation*}
Even though the fully-leptonic channel has a small cross section, it provides a clean measurement of the $ZZ$ final state.
The primary background comes from non-VBS diboson production (`SM ZZ QCD' in Fig.~\ref{fig:vbszz}).
A statistical analysis of the resulting $4\ell$ invariant mass distribution shown in Fig.~\ref{fig:vbszz} tests the hypothesis of the new ${\cal L}_{\phi W}$ operator against the null (SM) hypothesis.
The discovery significance for various values of the coefficient $ \frac{c_{\phi W}}{\Lambda^2}$ is also shown in Fig.~\ref{fig:vbszz}.
The $5\sigma$ discovery reach increases by more than a factor of two when the integrated luminosity changes from $300\,\ifb$ to $3000\,\ifb$.

\begin{figure}[htbp]
\begin{center}
\includegraphics[width=0.57\textwidth]{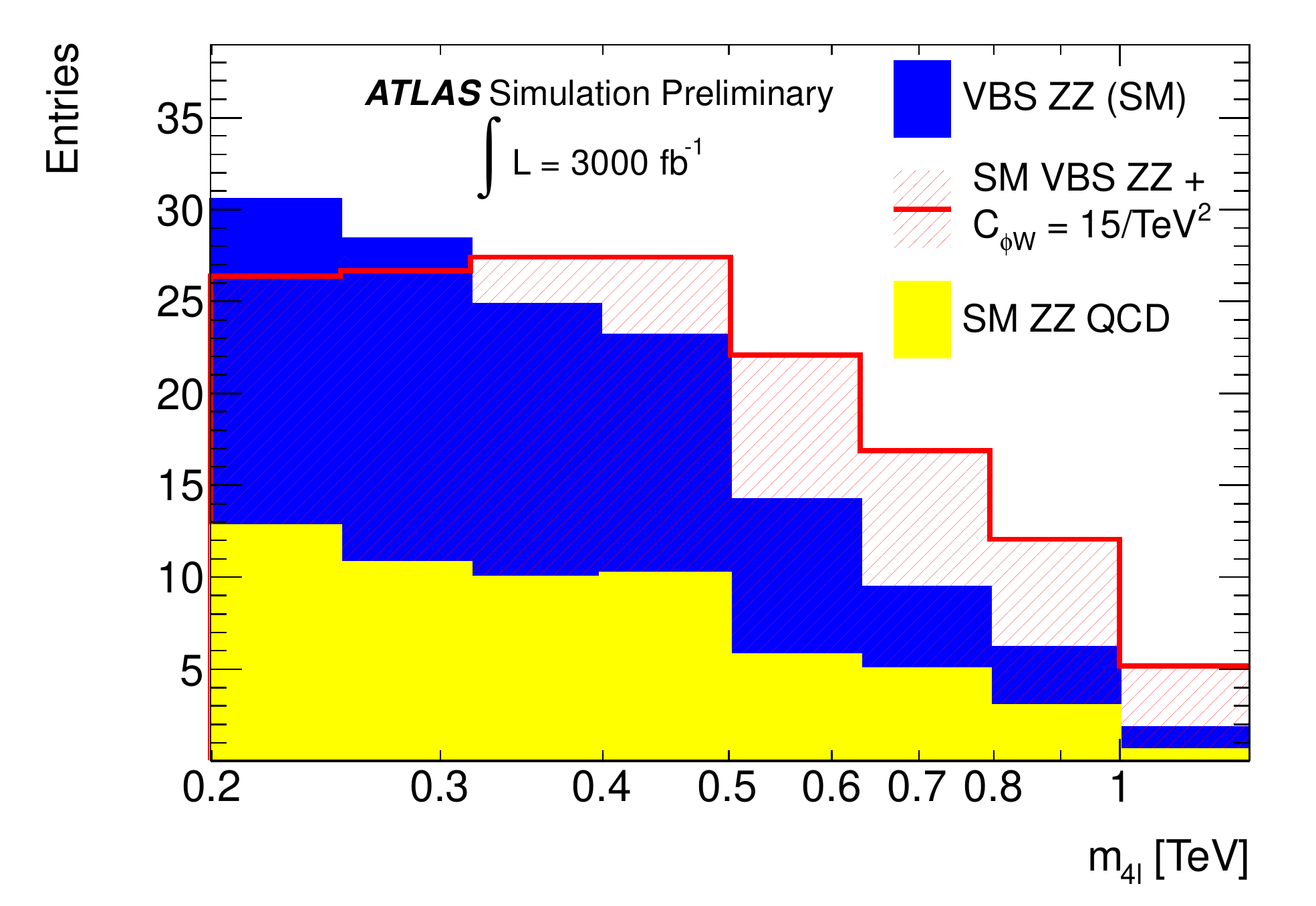}
\includegraphics[width=0.4\textwidth]{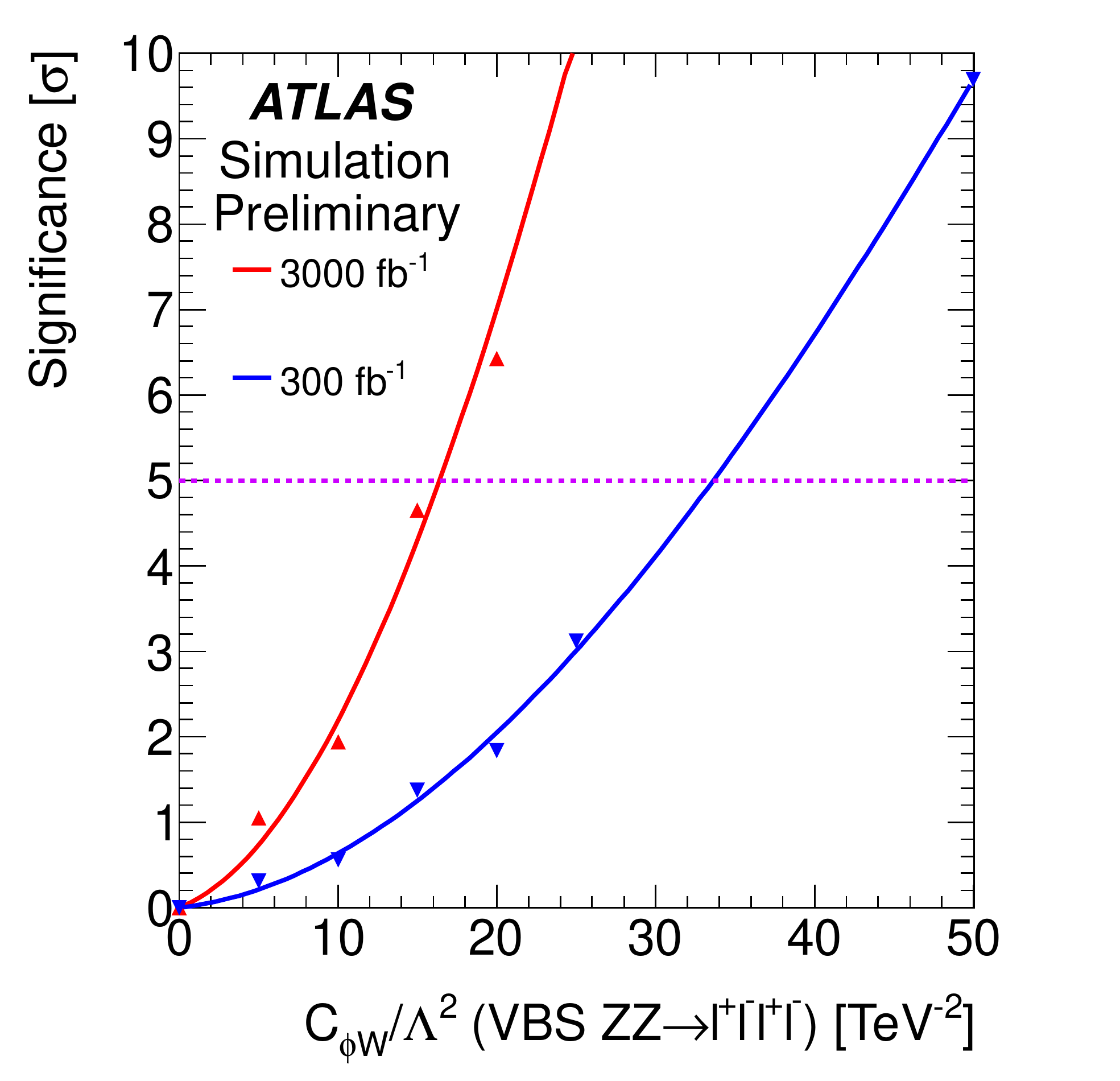}
\end{center}
\caption{Left: The reconstructed 4-lepton invariant mass distribution in $ZZ \to \ell\ell\ell\ell$ events.  Right: The signal significance as a function of $ \frac{c_{\phi W}}{\Lambda^2}$ (right).}
\label{fig:vbszz}
\end{figure}

Another potential vector boson scattering channel is the $WZ$ final state.
For this channel, the dimension-8 operator 
\begin{equation*}
    {\cal L}_{T,1} = \frac{f_{T1}}{\Lambda^4} {\rm Tr} [\hat{W}_{\alpha \nu} \hat{W}^{\mu \beta}] \times {\rm Tr} [\hat{W}_{\mu \beta}\hat{W}^{\alpha \nu}]
\end{equation*}
is chosen for study.  The $WZ$ final state benefits from a larger cross section than the $ZZ$ channel. The invariant mass can still be reconstructed in the fully leptonic channel by solving for the neutrino longitudinal momentum $p_Z$ under a $W$ mass constraint.
If all leptons have the same flavor, the lepton pair with invariant mass closest to $m_Z$ is taken to be the $Z$.
The sensitivity of this analysis to new physics is included in the summary of results in Table~\ref{tab:5sigewksummary}.

A third possible channel to investigate vector boson scattering is the same-sign $W^\pm W^\pm$ final state, and the dimension-8 operator
 \begin{equation*}
    {\cal L}_{S,0} = \frac{f_{S0}}{\Lambda^4} [(D_\mu \phi)^{\dagger} D_\nu \phi)] \times [(D^\mu \phi)^{\dagger} D^\nu \phi)].
 \end{equation*}
is used. Two selected leptons must have the same charge, and the invariant mass of the two highest-$p_T$ jets must be at least $1\,\text{TeV}$.  The primary backgrounds are Standard Model $WZ$ production, in which one of the leptons from the $Z$-decay is not identified,  and a small component of non-VBS $W^\pm W^\pm$ production (`SM ssWW QCD'). Misidentified-leptons, photon-conversions in $W\gamma$ events, and charge-flip contributions, collectively termed `mis-ID' backgrounds, were accounted for by scaling the $WZ$ background by a conservative factor of $\approx\!\! 2$, taken from a study of same-sign $WW$ production in the current ATLAS data.
The statistical analysis is performed by constructing templates of the $m_{lljj}$ distribution for different values of $f_{S0}/\Lambda^4$.
The distribution of $m_{lljj}$ and the signal significance as a function of $f_{S0}/\Lambda^4$ are shown in Fig.~\ref{fig:vbsssww}.

\begin{figure}[htbp]
\begin{center}
  \includegraphics[width=0.57\textwidth]{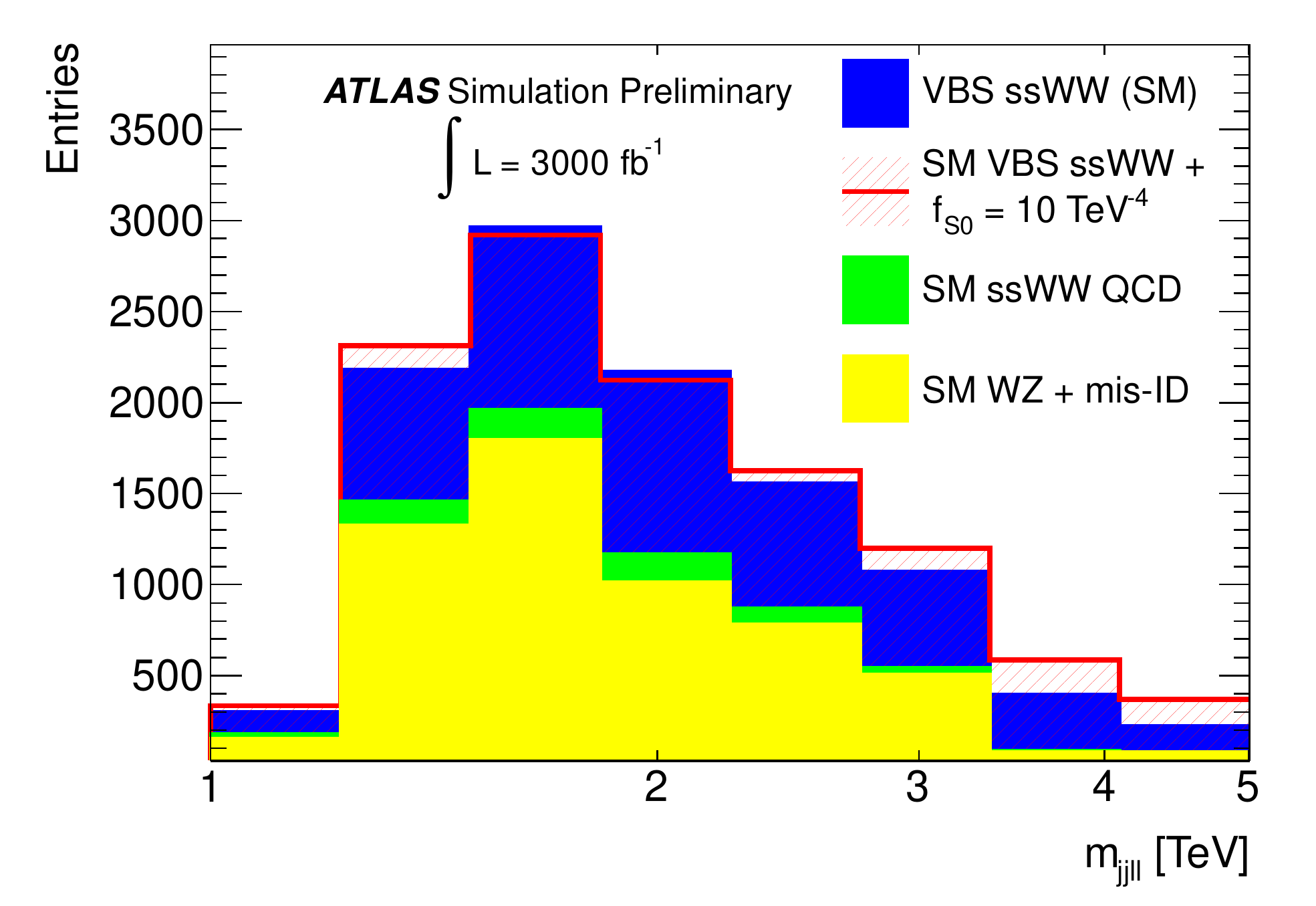}
  \includegraphics[width=0.4\textwidth]{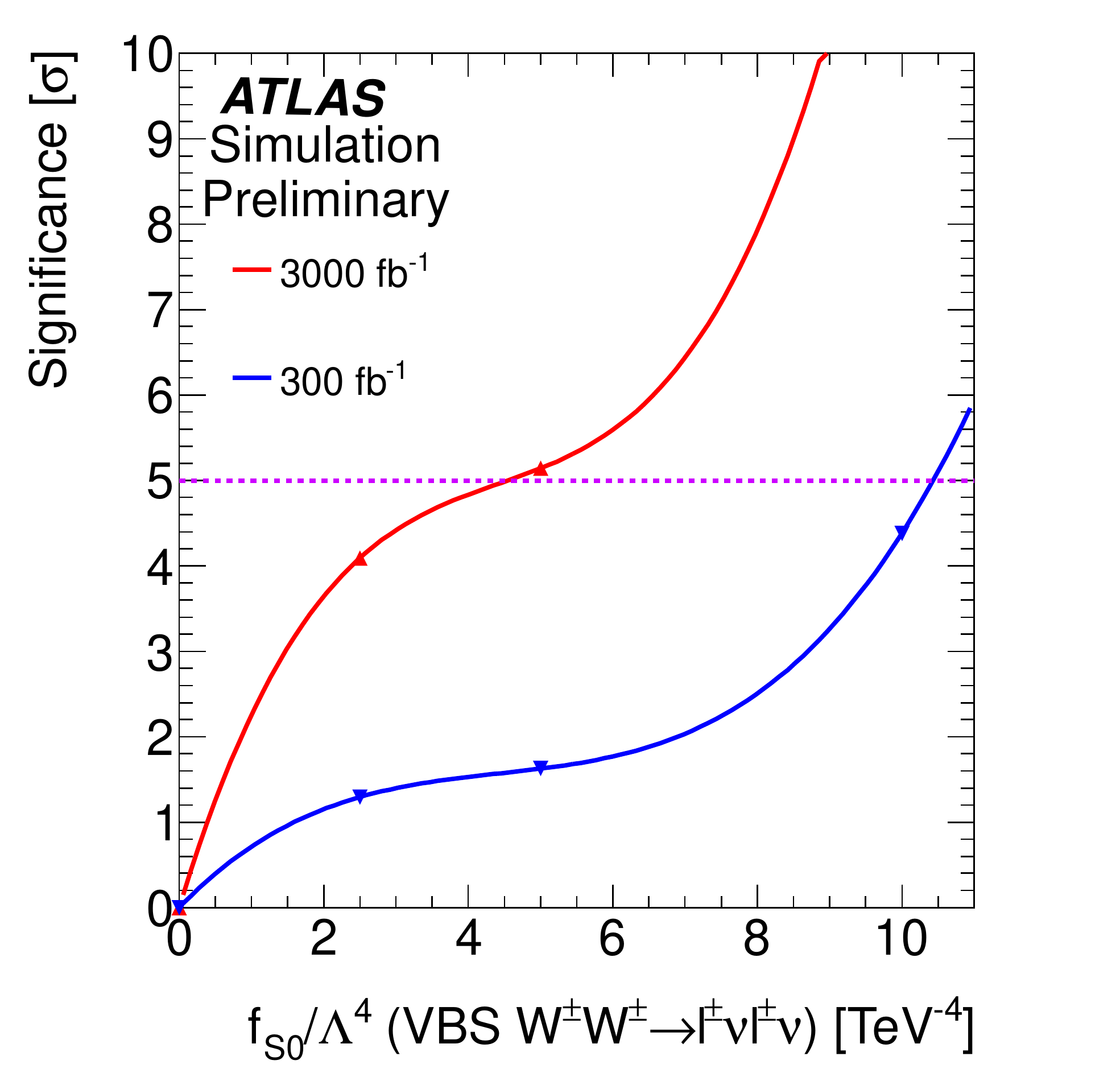}
\end{center}
\caption{Left: The reconstructed 4-body mass spectrum using the two leading leptons and jets, using the same-sign $WW \to \ell \nu \ell \nu$ VBS channel at $pp$ center-of-mass collision energy of 14 TeV. Right: The signal significance as a function of  $f_{S0}$.}
\label{fig:vbsssww}
\end{figure}

\subsection{Gauge Boson Couplings in Triboson Production}

The $Z\gamma\gamma$ mass spectrum at high mass is sensitive to BSM triboson contributions through quartic gauge couplings.
In this case, the lepton-photon channel allows full reconstruction of the final state and the $Z\gamma\gamma$ invariant mass.

Beyond the simple $Z$ reconstruction, additional requirements that $\Delta R(\ell,\gamma)>0.4$ and at least one $p_T (\gamma) > 160\,\GeV$ reduce the FSR contribution.  This restricts the measurement to a phase space that is uniquely sensitive to quartic gauge couplings (QGC).
The dominant process in the QGC-sensitive kinematic phase space is the Standard Model $Z\gamma\gamma$ production, while the backgrounds from $Z\gamma j$ and $Zjj$, with one or two jets misidentified as a photon, are subdominant.

The new BSM effective operators chosen to study triboson production are 
\begin{eqnarray}
\nonumber
{\cal L}_{T,8} & = & \frac{f_{T8}}{\Lambda^4} B_{\mu \nu}B^{\mu \nu} B_{\alpha \beta} B^{\alpha \beta} \nonumber \\
{\cal L}_{T,9} & = & \frac{f_{T9}}{\Lambda^4} B_{\alpha \mu}B^{\mu \beta} B_{\beta \nu}B^{\nu \alpha}.
\end{eqnarray}
which are uniquely probed by final states with neutral particles.
Fig.~\ref{fig:zgammagamma} shows the reconstructed $Z\gamma\gamma$ mass spectrum and expected discovery significance for the ${\cal L}_{T,8}$ dimension-8 electroweak operator.

\begin{figure}[htbp]
\begin{center}
\includegraphics[width=0.54\textwidth]{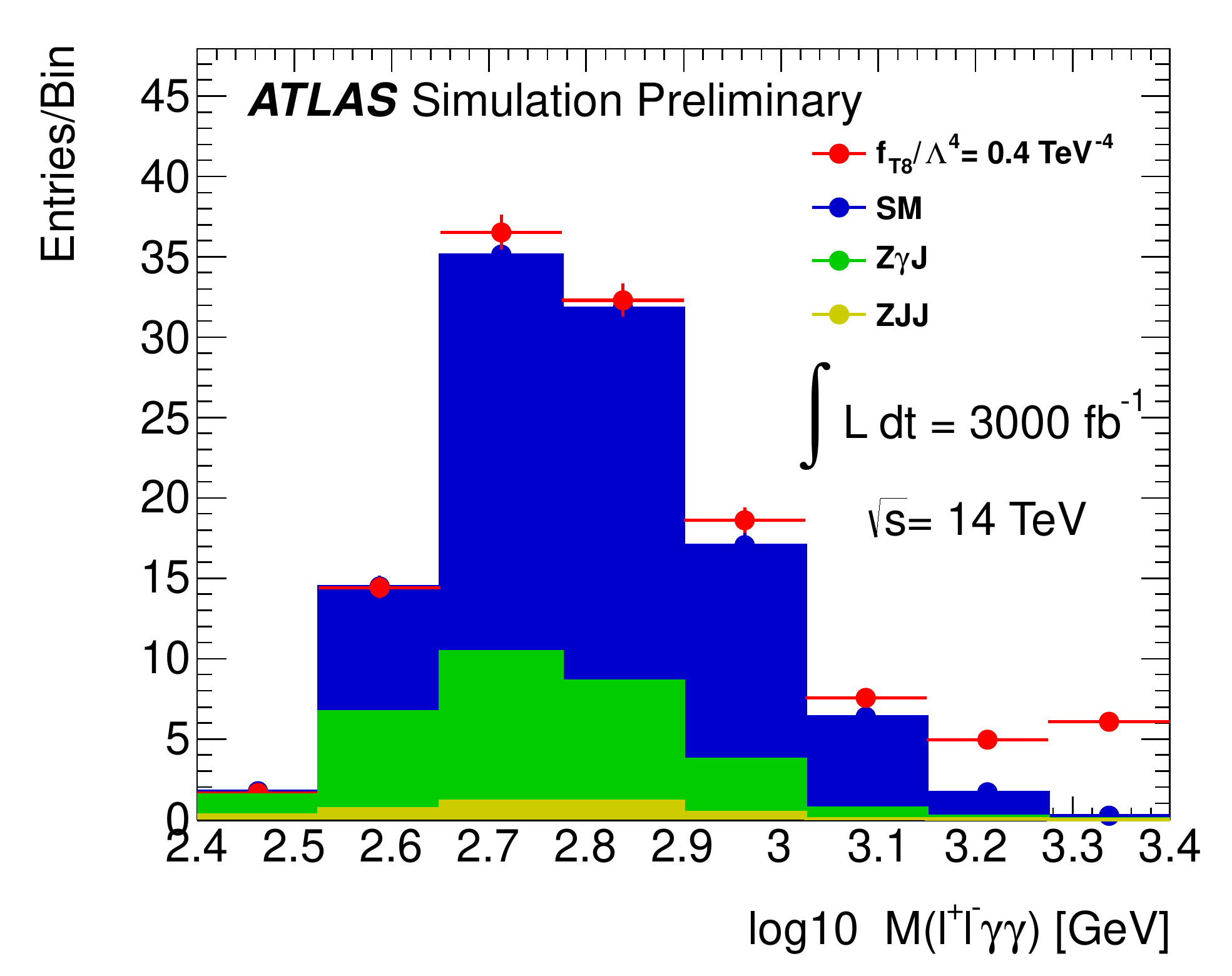}
\includegraphics[width=0.44\textwidth]{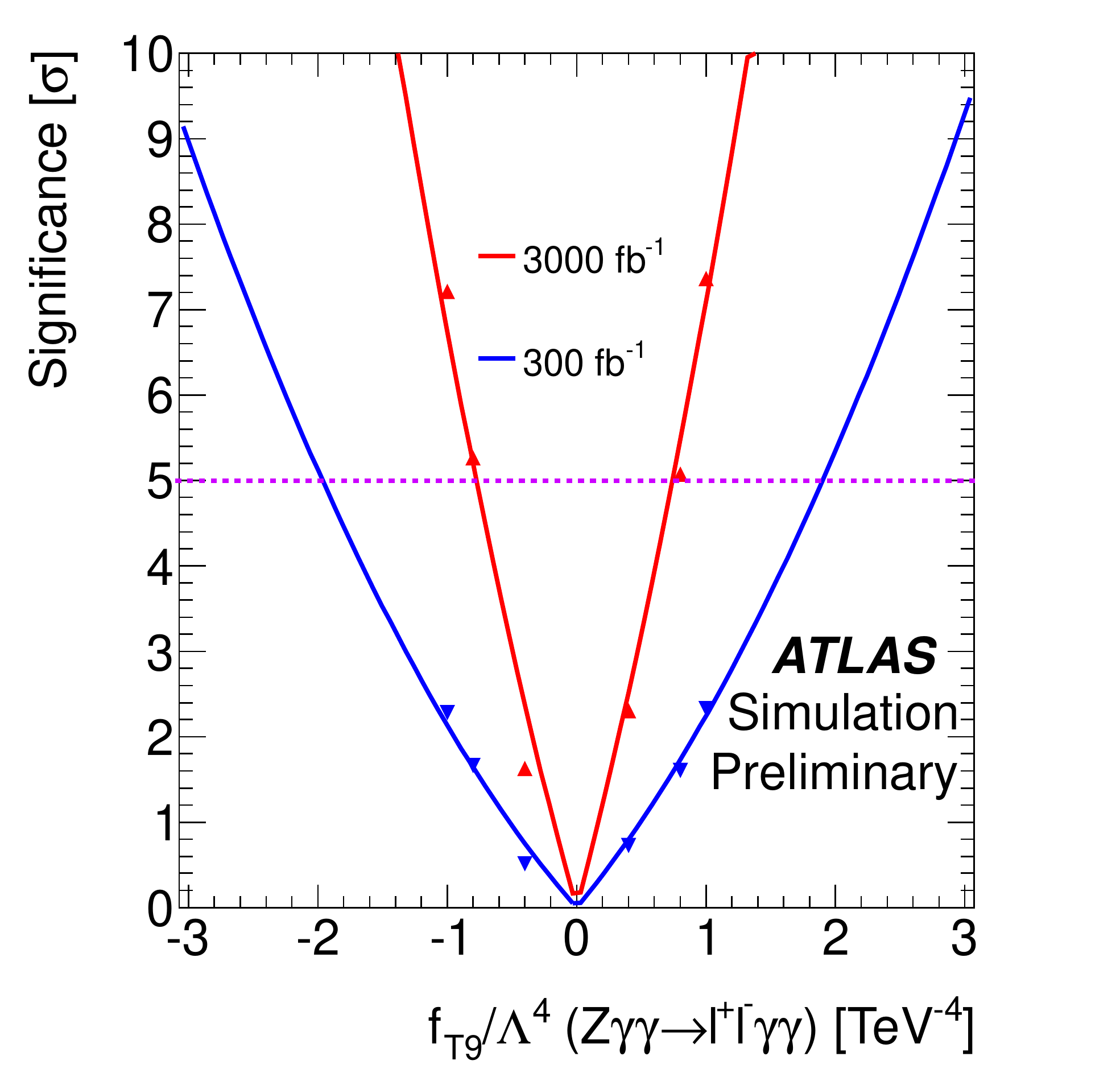}

\end{center}
\caption{Left: Reconstructed mass spectrum for the charged leptons and photons in selected $Z\gamma\gamma$ events. Right: The signal significance as a function of $f_{T9}/\Lambda^4$.}
\label{fig:zgammagamma}
\end{figure}
\subsection{Summary of Multiboson Studies}
The higher-luminosity HL-LHC dataset increases the discovery range for these new higher-dimension electroweak operators by more than a factor of two, as shown in Table~\ref{tab:5sigewksummary}.
If new physics in the electroweak sector is discovered in the $300\,\ifb$ dataset, then the coefficients on the new operators can be measured with 5\% precision in the $3000\,\ifb$ dataset.
\begin{table}[h]
\centering
\begin{tabular}{c|c|c|c|c|c|c|c}
\hline\hline
\multirow{2}{*}{Parameter} & \multirow{2}{*}{dimension} & \multirow{2}{*}{channel} & \multirow{2}{*}{$\Lambda_{UV}$ [TeV]} & \multicolumn{2}{|c|}{300 fb$^{-1}$} & \multicolumn{2}{|c}{3000 fb$^{-1}$}  \\
\cline{5-8}
                           &                            &                          &                                       & $5 \sigma$ & 95\% CL              & $5 \sigma$ & 95\% CL		     \\
\hline
        $c_{\phi W}/\Lambda^2$ & 6 & $ZZ$ & 1.9 & 34 TeV$^{-2}$ &  20 TeV$^{-2}$  & 16 TeV$^{-2}$  & 9.3 TeV$^{-2}$ \\
        $f_{S0}/\Lambda^{4}$ & 8 & $W^\pm W^\pm$ & 2.0 & 10 TeV$^{-4}$ & 6.8 TeV$^{-4}$  &  4.5 TeV$^{-4}$  & 0.8 TeV$^{-4}$ \\
        $f_{T1}/\Lambda^{4}$ & 8 & $WZ$ & 3.7 & 1.3 TeV$^{-4}$ & 0.7 TeV$^{-4}$ & 0.6 TeV$^{-4}$  & 0.3 TeV$^{-4}$ \\
        $f_{T8}/\Lambda^{4}$ & 8 & $Z\gamma\gamma$ & 12 & 0.9 TeV$^{-4}$ & 0.5 TeV$^{-4}$ & 0.4 TeV$^{-4}$  & 0.2 TeV$^{-4}$ \\
        $f_{T9}/\Lambda^{4}$ & 8 & $Z\gamma\gamma$ & 13 & 2.0 TeV$^{-4}$ & 0.9 TeV$^{-4}$ & 0.7 TeV$^{-4}$  & 0.3 TeV$^{-4}$ \\
\hline\hline
\end{tabular}
\caption{$5 \sigma$-significance discovery values and 95\% CL limits for coefficients of higher-dimension electroweak operators.  $\Lambda_{UV}$ is the unitarity violation bound corresponding to the sensitivity with 3000 fb$^{-1}$ of integrated luminosity.
 }
\label{tab:5sigewksummary}
\end{table}

%% file: susy.tex
\section{Searches for New Particles Predicted by Theories of Supersymmetry}

Supersymmetry (SUSY) is an extended symmetry relating fermions and bosons.
In theories of supersymmetry, every SM boson (fermion) has a supersymmetric fermion (boson) partner.
Extending the sensitivity of the ATLAS experiment to these new particles is one of the key aspects of the HL-LHC physics program.

In $R$-parity conserving supersymmetric extensions of the SM, SUSY particles are produced in pairs, either through strong or weak production, and these particles decay in a cascade of SUSY and SM particles.
The lightest supersymmetric particle (LSP) is stable in these $R$-parity conserving extensions.
As a result, the searches for evidence of SUSY particle production focus on experimental signatures with large missing transverse momentum from undetected LSPs.

A high-luminosity dataset benefits especially the searches for particles produced in small cross section interactions or in signatures with small branching fractions.
Three representative searches and their potential for discovery with a $3000\,\ifb$ dataset are presented in the following subsections~\cite{ATL-PHYS-PUB-2013-002}.
These results are only indicative of future discovery prospects, and in fact are understood to be fairly conservative, since they depend on conservative performance assumptions and analysis strategies.

\subsection{Direct Production of Weak Gauginos}

Weak gauginos can be produced in decays of squarks and gluinos or directly in weak production.
For weak gaugino masses of several hundred GeV, as expected from naturalness arguments~\cite{Papucci:2011wy}, the weak production cross section is rather small, ranging from $10^{-2}$ to $10\,\text{pb}$, and a dataset corresponding to high integrated luminosity is necessary to achieve sensitivity to high-mass weak gaugino production.
Results with the 2012 data exclude charginos masses of 300 to $600\,\text{GeV}$ for small LSP masses, depending on whether sleptons are present in the decay chain.
For LSP masses greater than $100\,\text{GeV}$ there are currently no constraints from the LHC if the sleptons are heavy .

The weak gauginos can decay via $\tilde{\chi}_2^0 \rightarrow Z \tilde{\chi}_1^0$ or $\tilde{\chi}^\pm_1 \rightarrow W^\pm \tilde{\chi}^0_1$, and both of these decays lead to a final state with three leptons and large missing transverse momentum.
SM background for this final state is dominated by the irreducible $WZ$ process, even with a high missing transverse momentum requirement of $150\,\text{GeV}$.
Boosted decision trees can be trained to use kinematic variables, such as the leptons$^\prime$ transverse momenta, the $p_T$ of the Z-boson candidate, the summed $E_T$ in the event, and the transverse mass $m_T$ of the lepton from the $W$ and the missing transverse momentum.

The expected sensitivity for the search is calculated using a simplified model in which the $\tilde{\chi}^0_2$ and $\tilde{\chi}^\pm_1$ are nearly degenerate in mass.
With a ten-fold increase in integrated luminosity from 300 to $3000\,\ifb$, the discovery reach extends to chargino masses above $800\,\GeV$, to be compared with the reach of $350\,\GeV$ from the smaller dataset.
The extended discovery reach and comparison are shown in Fig.~\ref{fig:ewkgauginos}.

\begin{figure}[htbp]
\begin{center}
\includegraphics[width=0.8\textwidth]{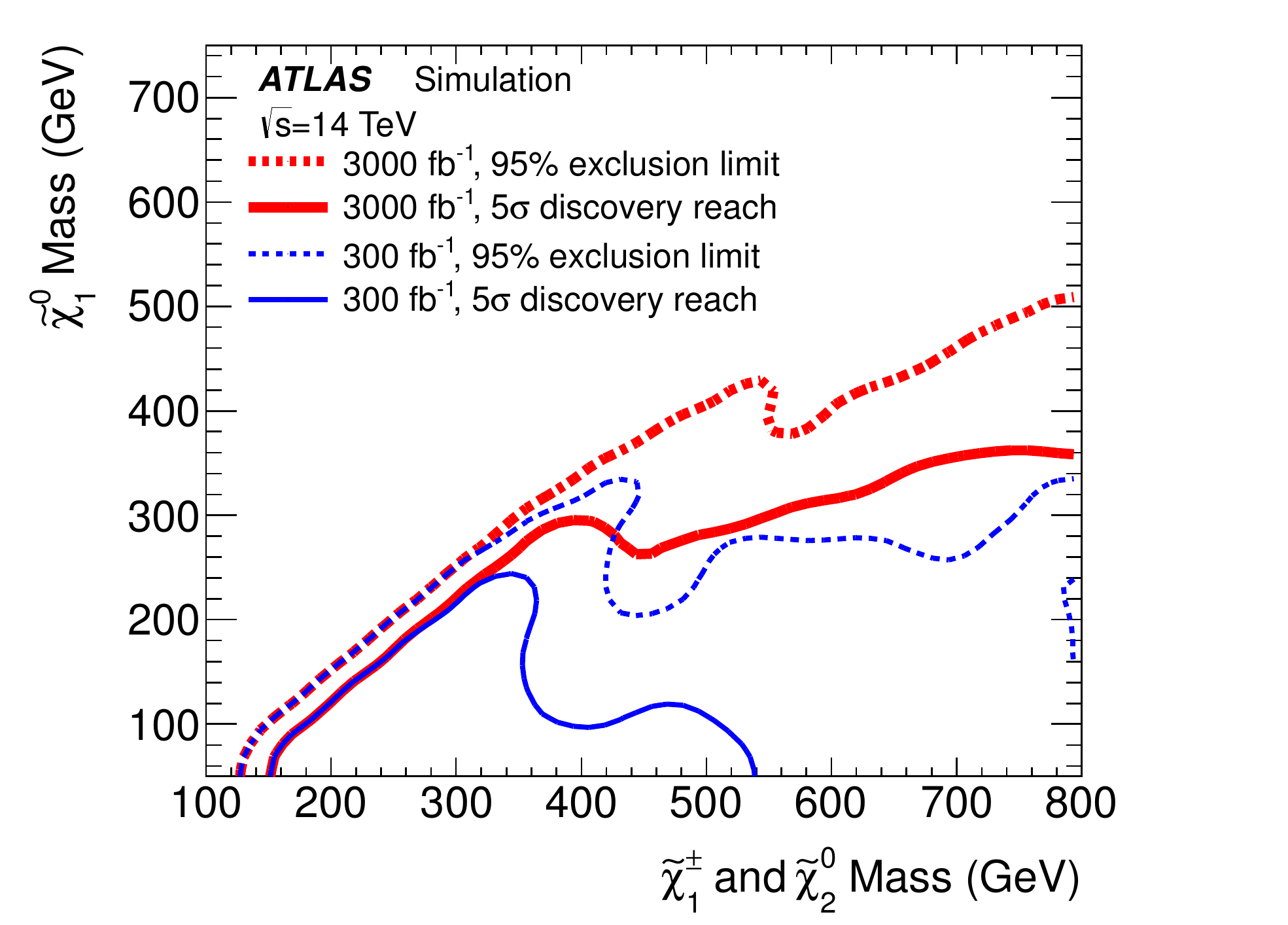}
\end{center}
\caption{Discovery reach (solid lines) and exclusion limits (dashed lines) for charginos and neutralinos in $\tilde{\chi}^\pm_1 \tilde{\chi}_2^0 \rightarrow W^{(\star)} \tilde{\chi}_1^0 Z^{(\star)} \tilde{\chi}^0_1$ decays.
The results are shown for the $300\,\text{fb}^{-1}$ and $3000\,\text{fb}^{-1}$ datasets.}
\label{fig:ewkgauginos}
\end{figure}

\subsection{Direct Production of Top Squarks}

Naturalness arguments lead to the conclusion that a Higgs boson mass of $m_H=125\,\GeV$ favors a light top squark mass, less than $1\,\TeV$.
A direct search for top squarks needs to cover this allowed range of masses.
The top squark pair production cross section at $\sqrt{s}=14\,\TeV$ is $10\,\text{fb}$ for $m_{\tilde{t}}=1\,\TeV$.
For the purpose of this study, the stops are assumed to decay either to a top quark and the LSP ($\tilde{t} \rightarrow t + \tilde{\chi}^0_1$) or to a bottom quark and the lightest chargino ($\tilde{t} \rightarrow b + \tilde{\chi}^\pm_1$).
The final state for the first decay is a top quark pair in associated with large missing transverse momentum, while the final state for the second decay is 2 $b$-jets, 2 $W$ bosons, and large missing transverse momentum.
In both cases, leptonic signatures are used to identify the top quarks or the $W$ bosons.
The 1-lepton + jet channel is sensitive to $\tilde{t} \rightarrow t + \tilde{\chi}^0_1$, and the 2-lepton + jet channel is sensitive to $\tilde{t} \rightarrow b + \tilde{\chi}_1^\pm$.
For this study, the event selection requirements were not reoptimized for a greater integrated luminosity.

An increase in the integrated luminosity from 300 to $3000\,\ifb$ results in an increase in a stop mass discovery reach of approximately 150 GeV, up to $920\,\text{GeV}$ (see Fig.~\ref{fig:stopreach}).
This increase covers a significant part of the top squark range favored by naturalness arguments.  In this study the same selection cuts were used for the two luminosity values. 

\begin{figure}[htbp]
\begin{center}
\includegraphics[width=0.8\textwidth]{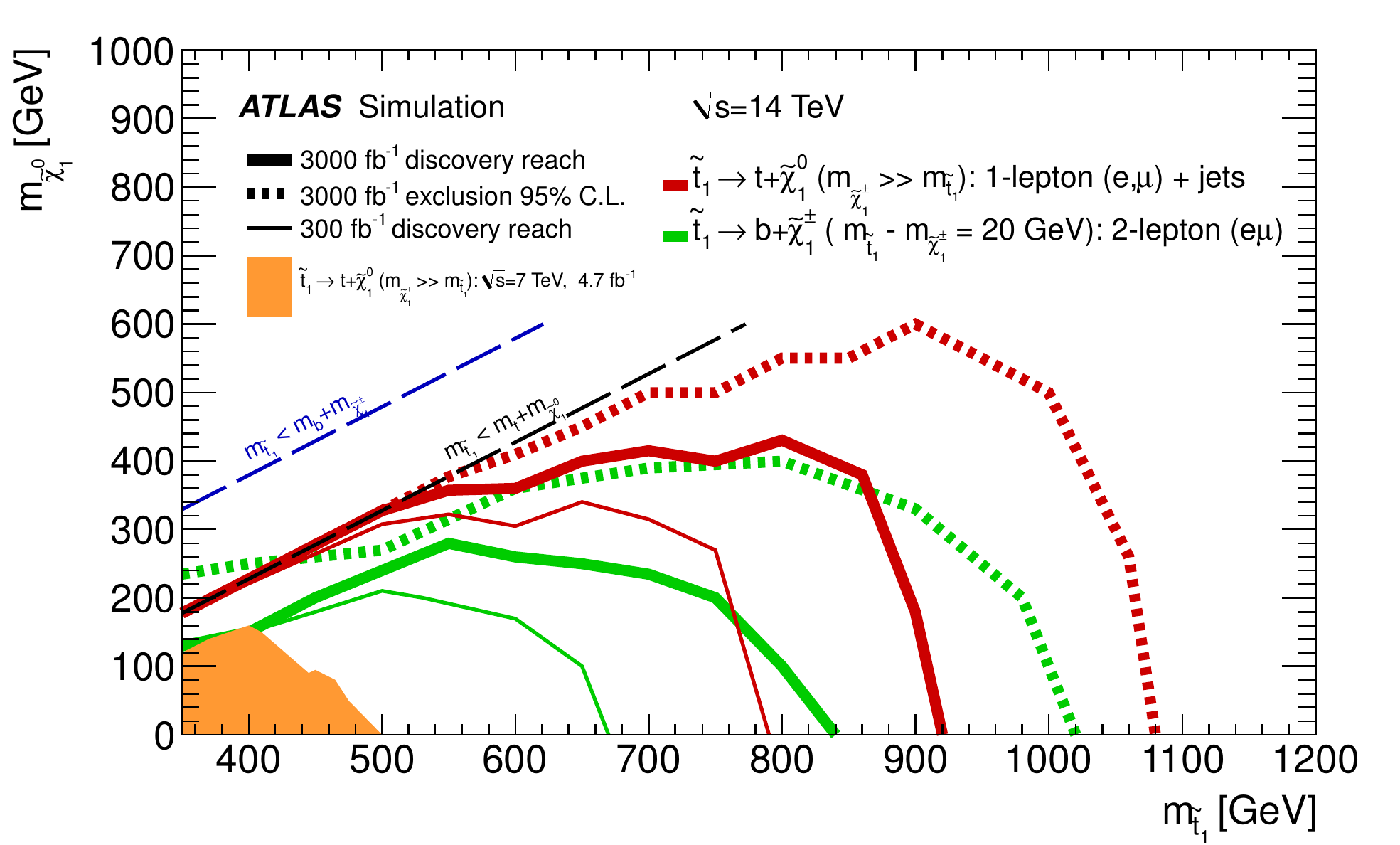}
\end{center}
\caption{Discovery reach (solid lines) and exclusion limits (dashed lines) for top squarks in the $\tilde{t} \rightarrow t+\tilde{\chi}^0_1$ (red) and the $\tilde{t} \rightarrow b+\tilde{\chi}^\pm_1, \tilde{\chi}^\pm_1 \rightarrow W+\tilde{\chi}_1^0$ (green) decay modes.}
\label{fig:stopreach}
\end{figure}

\subsection{Strong Production of Squarks and Gluinos}

A high-luminosity dataset would allow the discovery reach for gluinos and squarks to be pushed to the highest masses.
Gluinos and light-flavor squarks can be produced with a large cross section at $14\,\text{TeV}$, and the most striking signature is still large missing transverse momentum as part of large total effective mass.
An optimized event selection for a benchmark point with $m_{\tilde{q}} = m_{\tilde{g}} = 3200\,\text{GeV}$ requires the missing transverse momentum significance, defined as $E_T^\text{miss}/\sqrt{H_T}$, be greater than $15\,\text{GeV}^{1/2}$.
(The variable $H_T$ is defined to be the scalar sum of the jet and lepton transverse energies and the missing transverse momentum in the event.)
Both the missing $E_T$ significance and the effective mass are shown for the representative points in Fig.~\ref{fig:susymet}.

\begin{figure}[htbp]
\begin{center}
\includegraphics[width=0.45\textwidth]{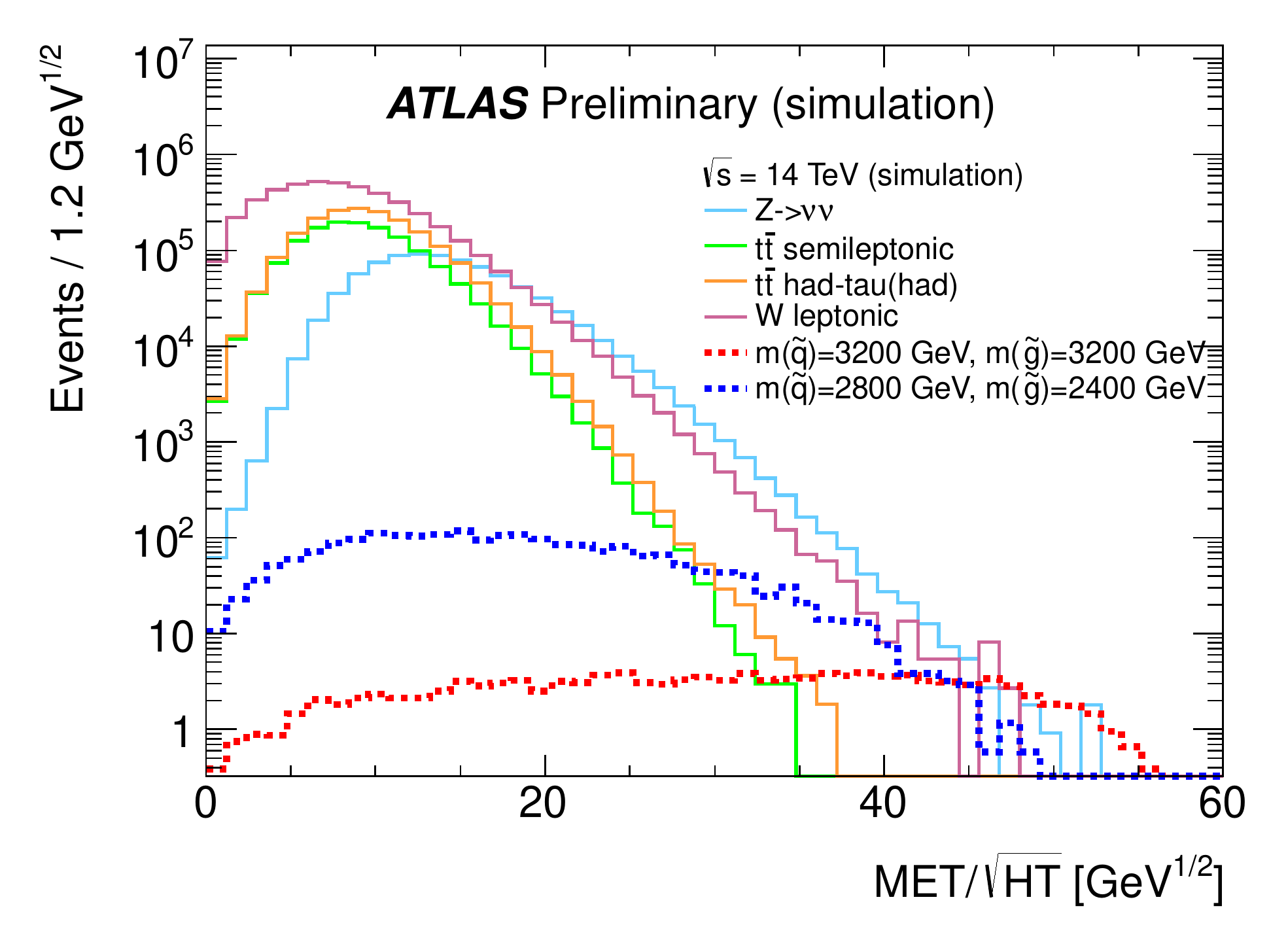}
\includegraphics[width=0.45\textwidth]{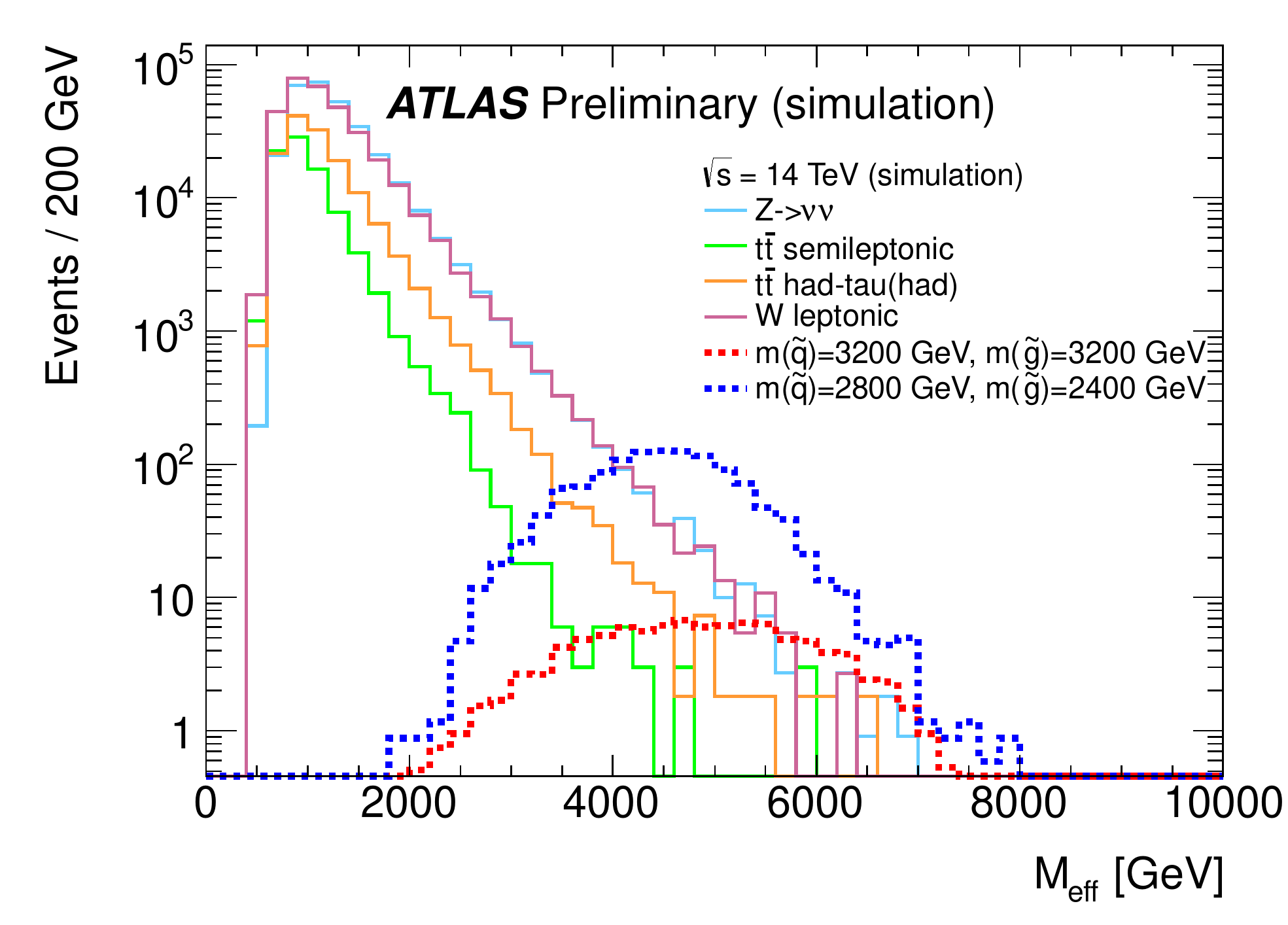}
\end{center}
\caption{Distribution of missing $E_T$ significance for SM backgrounds and two example SUSY benchmark points, normalized to 3000~\ifb (left), and
distributions of the effective mass (right), also normalized to 3000~\ifb.
The events shown in the effective mass distribution have passed the missing $E_T$ significance $15$~\GeV$^{1/2}$ requirement, the lepton veto, and the jet multiplicity requirement (at least 4 jets with $p_T > 60\,\text{GeV}$).}
\label{fig:susymet}
\end{figure}

The simple cut requirements on $H_T$, $M_\text{eff}$ and the $E_T^\text{miss}$ significance are re-optimized for the high-luminosity dataset of $3000\,\ifb$.
An increase in integrated luminosity from 300 to $3000\,\ifb$ results in a 400 GeV increase in the discovery reach, as shown in Fig.~\ref{fig:strongreach}.

\begin{figure}[htbp]
\begin{center}
\includegraphics[width=0.8\textwidth]{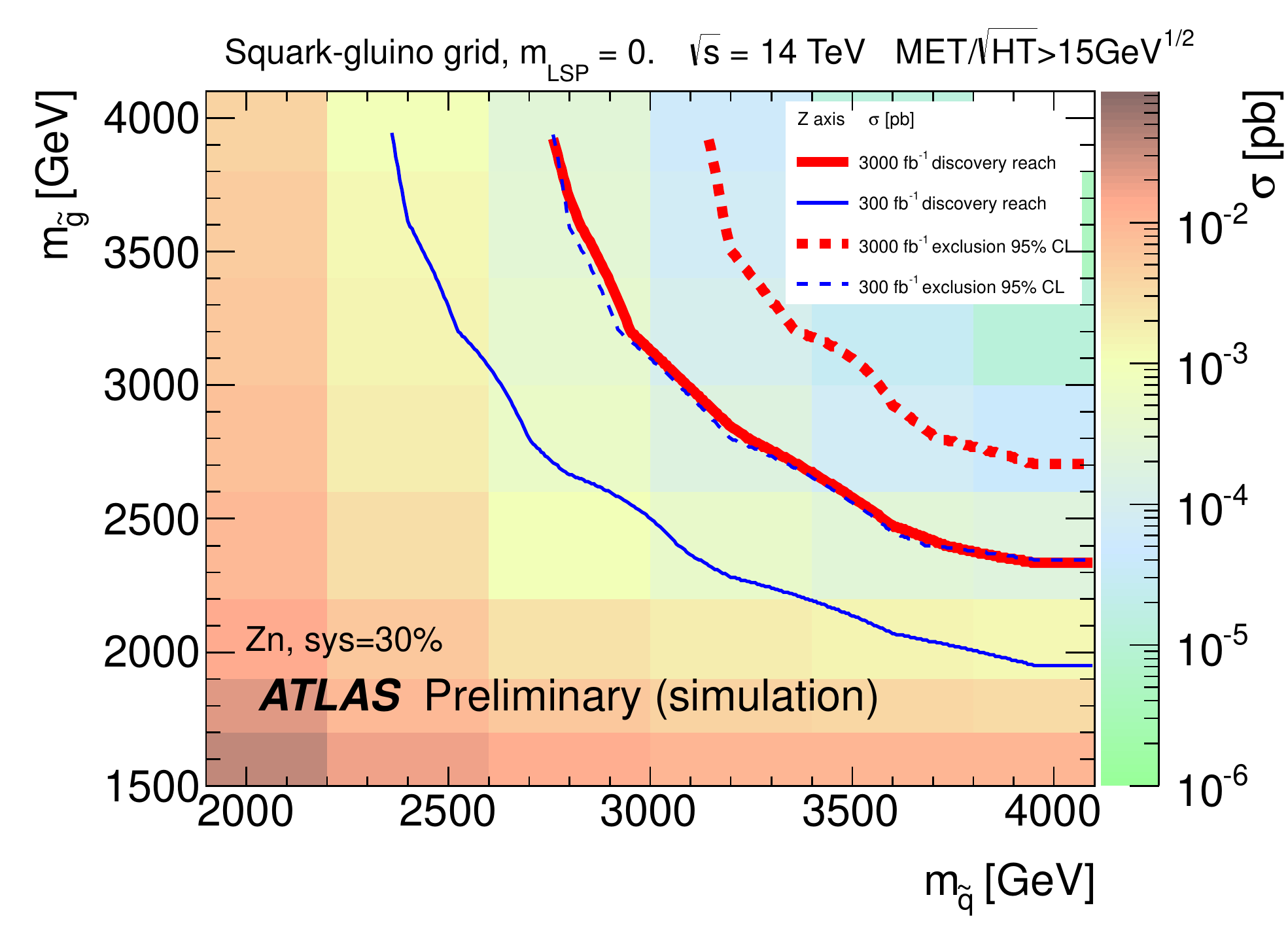}
\end{center}
\caption{Discovery reach and 95\% CL limits in a simplified squark--gluino model with a massless neutralino. The color scale shows the $\sqrt{s}=14\,\TeV$ NLO cross-section. The solid (dashed) lines show the $5\sigma$ discovery reach (95\% CL exclusion limit) with 300~\ifb\ and with 3000~\ifb , respectively.}
\label{fig:strongreach}
\end{figure}

%% file: exotics.tex
\section{Searches for Exotic Particles and Interactions}
The HL-LHC substantially increases the potential for the discovery of exotic new phenomena.  The range of possible phenomena is quite large.  In this section we discuss two benchmark exotic models of BSM physics and the expected gain in sensitivity from the order of magnitude increase in integrated luminosity provided by the HL-LHC.

\subsection{Searches for $t\bar t$ Resonances}

Strongly- and weakly-produced $\ttbar$ resonances provide benchmarks
not only for cascade decays containing leptons, jets (including
$b$-quark jets) and \met, but also the opportunity to study
highly boosted topologies. The sensitivity to the Kaluza-Klein gluon ($g_{KK}$) via the process $pp \to g_{KK} \to \ttbar\,$ and a heavy $Z^\prime$ decaying to $\ttbar$ at the HL-LHC is studied in both the dilepton and the lepton+jets decay
modes of the $\ttbar$ pair~\cite{PUB-2013-003}.

\begin{figure}[h]
  \centering
  \includegraphics[width=0.45\textwidth]{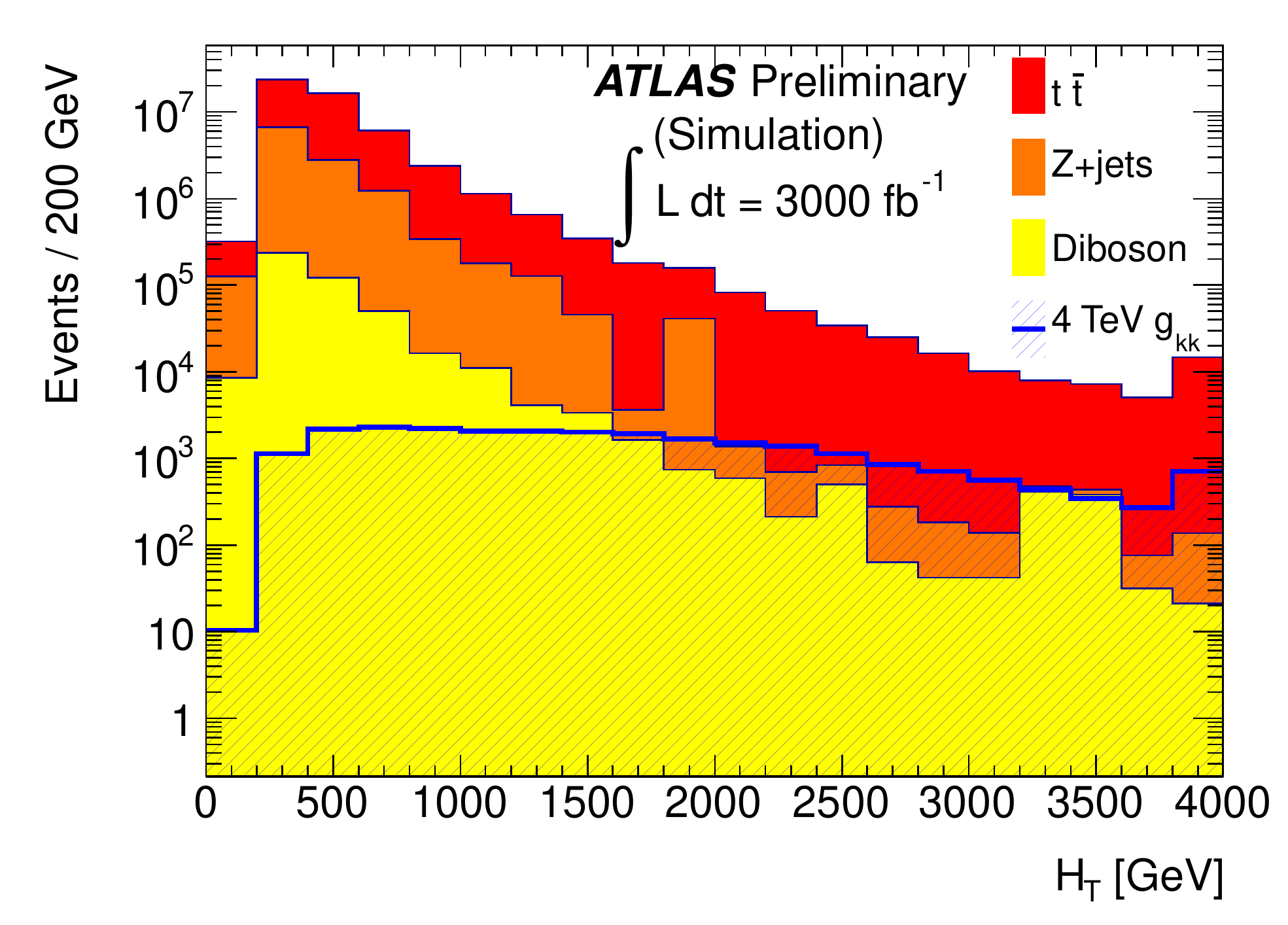}
  \includegraphics[width=0.45\textwidth]{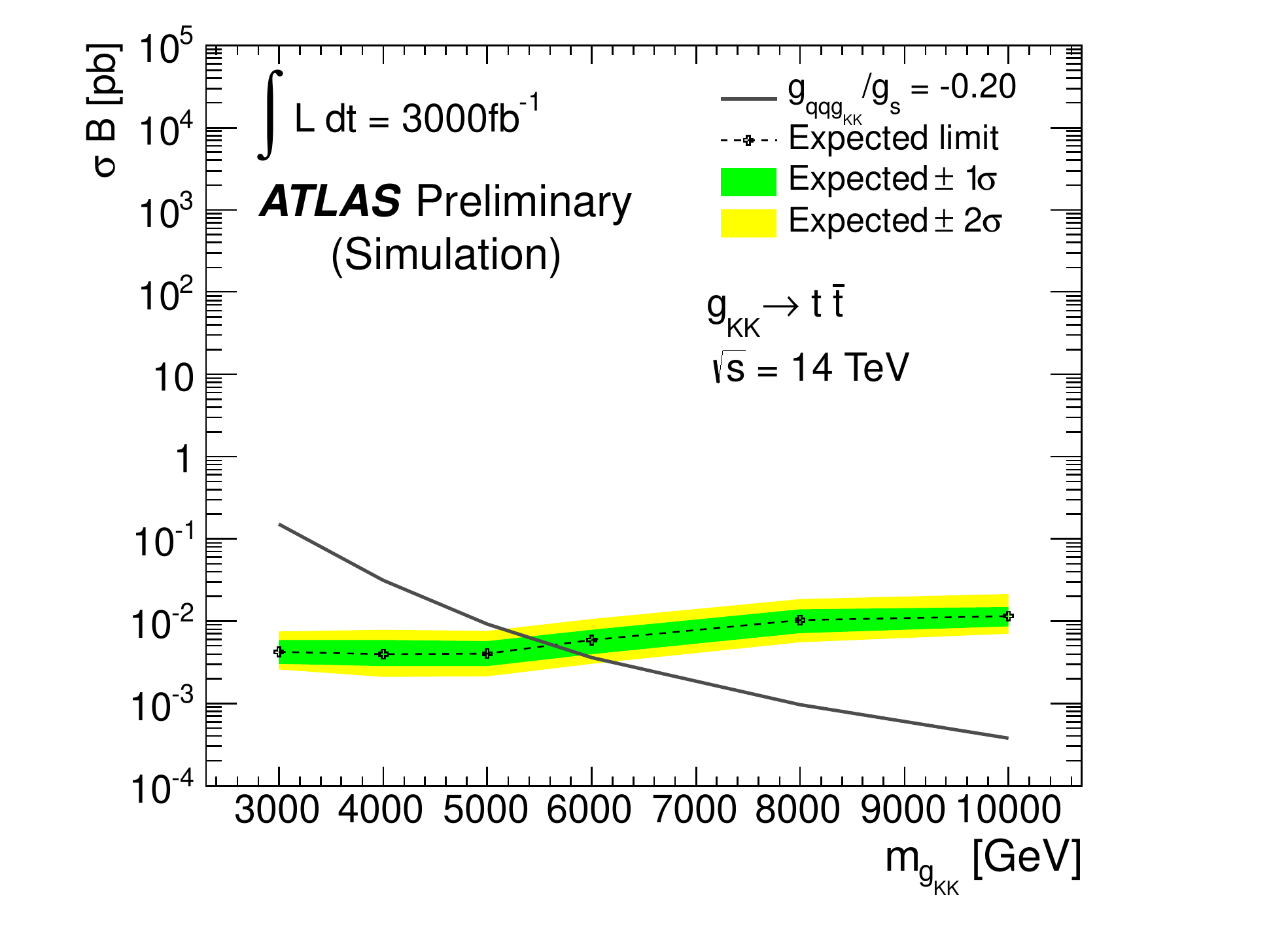}
  \caption{\label{fig:kkgluondilep} Left: The reconstructed resonance $H_{T}$
  spectrum for the $g_{KK} \to \ttbar$
  search in the dilepton channel with $3000\,\ifb$ for $pp$ collisions
  at $\sqrt{s} =14 \TeV$.
  The highest-$H_{T}$ bin includes the overflow. Right: The 95\% CL limit on the cross section times branching ratio.  Also shown is the theoretical expectation for the $g_{KK}$ cross section, for a ratio of the coupling to quarks to $g_s$ of -0.2, where $g_s=\sqrt{4\pi\alpha_s}$. }
\end{figure}

\begin{figure}[h]
  \centering
\includegraphics[width=0.45\textwidth]{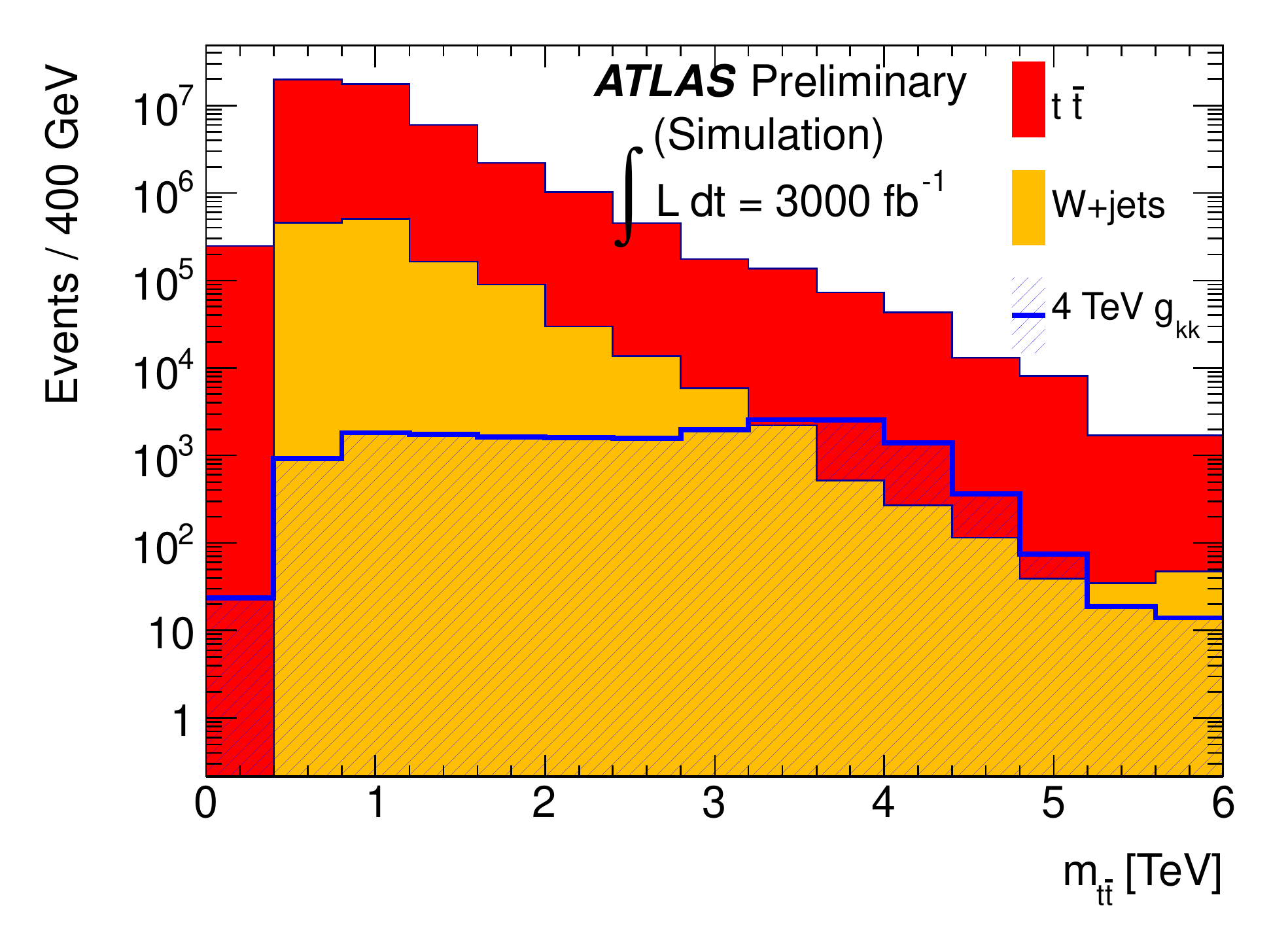}
\includegraphics[width=0.45\textwidth]{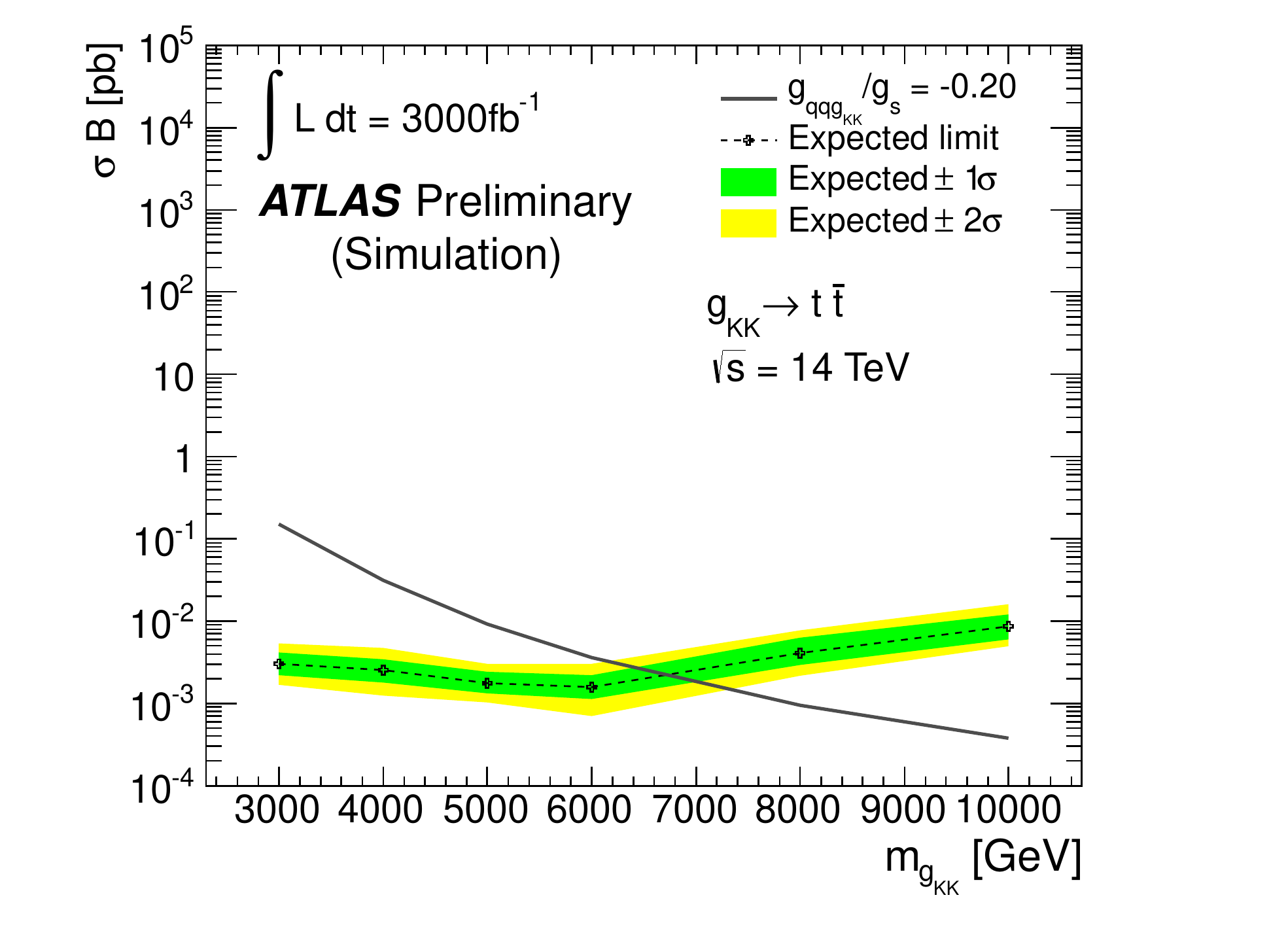}
  \caption{\label{fig:kkgluonljets} Left: The reconstructed resonance mass
  spectrum for the $g_{KK} \to
  \ttbar$ search in the lepton+jets channel with $3000\,\ifb$ for
  $pp$ collisions at $\sqrt{s} =14 \TeV$.
  The highest-mass bin includes the overflow. Right: The 95\% CL limit on the cross section times branching ratio.  Also shown is the theoretical expectation for the $g_{KK}$ cross section, for a ratio of the coupling to quarks to $g_s$ of -0.2, where $g_s=\sqrt{4\pi\alpha_s}$. }
\end{figure}

The two $\ttbar$ decay modes are complementary in that the lepton+jets mode allows a more complete reconstruction of the $\ttbar$ invariant mass, but suffers from more background, whereas the dilepton channel benefits from a smaller background contribution, but a more difficult reconstruction of the $\ttbar$ invariant mass.  In addition, in the case of boosted $\ttbar$ pairs, the dilepton decay mode is less affected by the merging of top quark decay products since the leptons are easier to identify close to a $b-$jet than are jets from the $W$ decay.  The lepton+jets mode therefore uses the reconstructed $\ttbar$ invariant mass distribution, while the dilepton mode uses the distribution of the scalar sum, $H_T$, of the $E_T$ of the two leading leptons, two leading jets, and missing $E_T$.  The statistical analysis is performed by a likelihood fit of templates of these distributions, using background plus varying amounts of signal, to the simulated data.
The $H_T$ and $m_{\ttbar}$ distributions and the resulting limits as a
function of the $g_{KK}$ pole mass for the dilepton and lepton+jets
channel are shown in Fig.~\ref{fig:kkgluondilep}
and Fig.~\ref{fig:kkgluonljets}, respectively.

The 95\% CL expected limits in the absence of signal, using statistical errors only, are shown in Table~\ref{tab:ttbar}.  The increase of a factor of ten in integrated luminosity, from 300 to 3000 fb$^{-1}$ raises the sensitivity to high-mass $\ttbar$ resonances by up to 2.4 TeV.

\begin{table}[h!]
\centering
\begin{tabular}{lccc}
\hline
model              & $300\,\ifb$  & $1000\,\ifb$     & $3000\,\ifb$  \\
\hline
\hline
$g_{KK}$ & 4.3 (4.0) & 5.6 (4.9)  & 6.7 (5.6)  \\
$Z^\prime_{\rm topcolor}$ & 3.3 (1.8) & 4.5 (2.6)  & 5.5 (3.2)  \\    
\hline
\end{tabular}

\caption{Summary of the expected limits for $g_{KK} \to \ttbar$  and
    $Z^\prime_{\rm topcolor} \to \ttbar$ searches in the lepton+jets
    (dilepton) channel for $pp$ collisions at $\sqrt{s} = 14 \TeV$.
    All limits are quoted in TeV.}
\label{tab:ttbar}

\end{table}

\subsection{Searches for Dilepton Resonances}

For studies of the sensitivity to a $Z^\prime$ boson~\cite{PUB-2013-003},
the dielectron and dimuon channels are considered separately since their momentum resolutions
scale differently with $p_T$ and the detector acceptances are
different. The background is dominated by the SM Drell-Yan production, while
$\ttbar$ and diboson backgrounds are substantially smaller.
Therefore, only the Drell-Yan background is considered in this study.
There is an additional background from non-prompt electrons due to
photon conversions which needs to be suppressed in the dielectron
channel.  The required rejection of this background is assumed to be
achieved with the upgraded detector.

\begin{figure}[h!]
  \centering
  \includegraphics[width=0.45\textwidth]{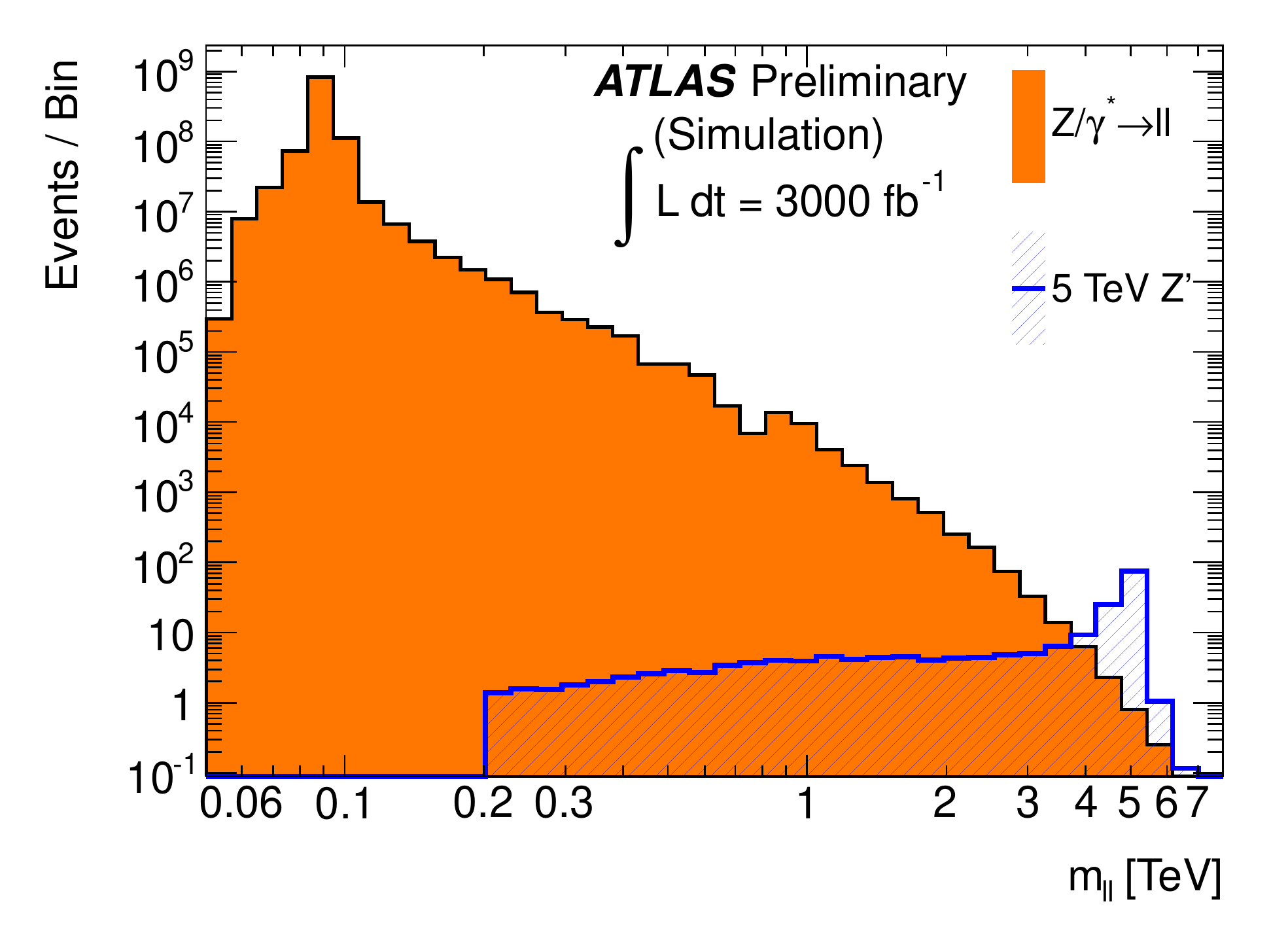}
  \includegraphics[width=0.45\textwidth]{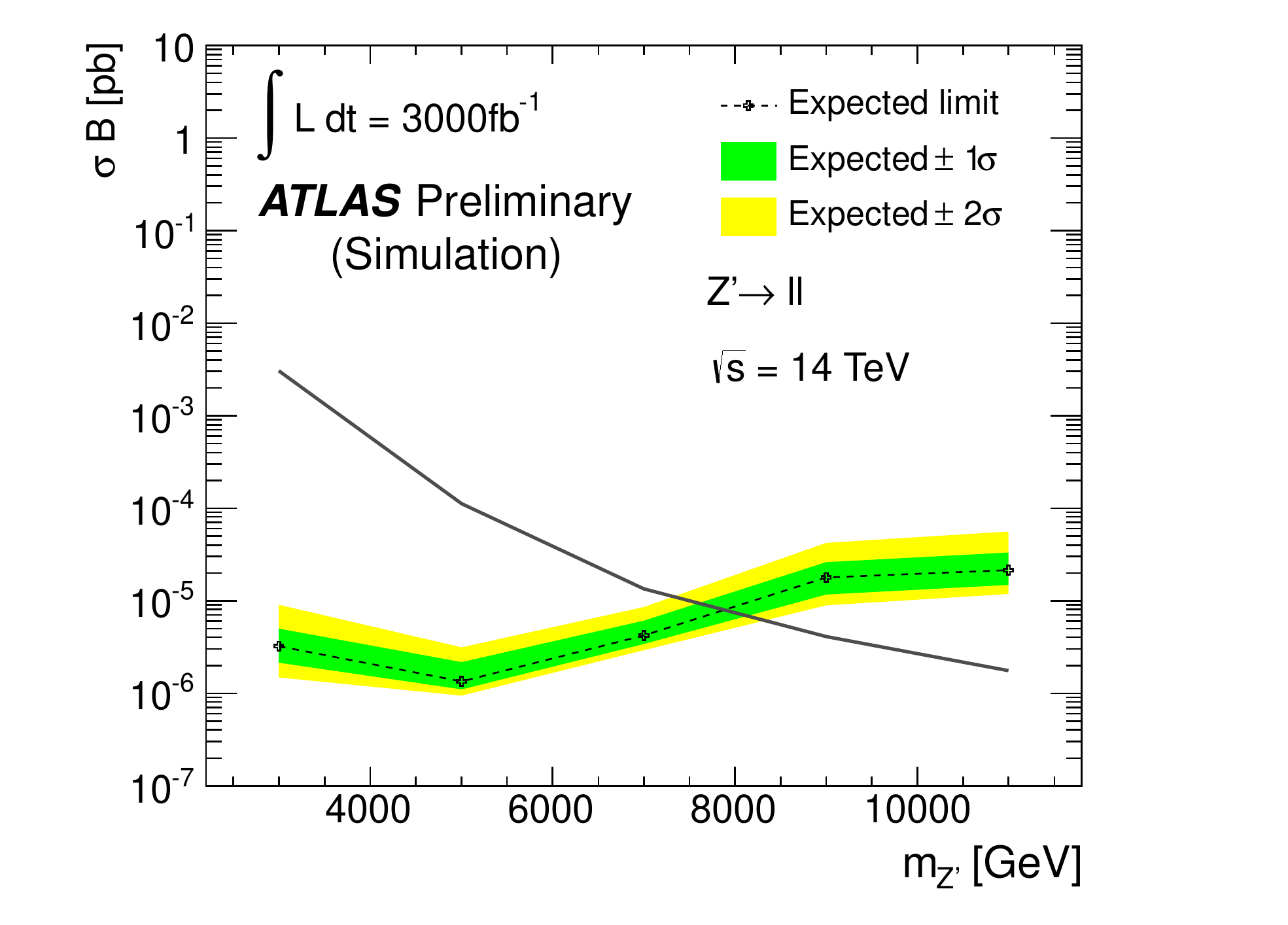}
  \caption{\label{fig:zprimeel} Left: The reconstructed dielectron mass
  spectrum for the $Z^\prime$ search with $3000\,\ifb$ of $pp$ collisions at $\sqrt{s} =14 \TeV$. The highest-mass bin includes the overflow. Right: The 95\% CL upper limit on the cross section times branching ratio. Also shown is the theoretical expectation for the $Z^\prime_{SSM}$ cross section.}
\end{figure}

\begin{figure}[h!]
  \centering
  \includegraphics[width=0.45\textwidth]{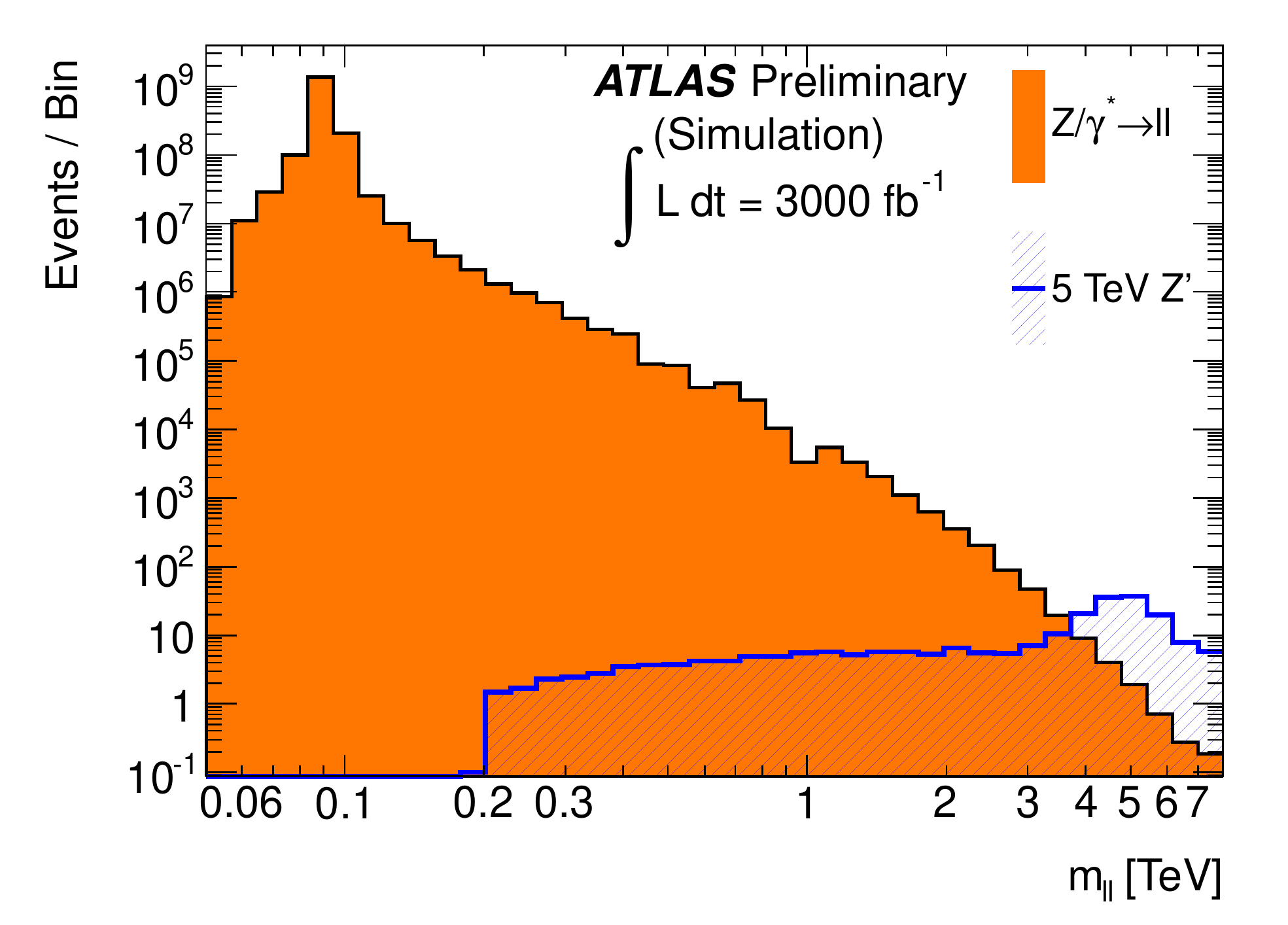}
  \includegraphics[width=0.45\textwidth]{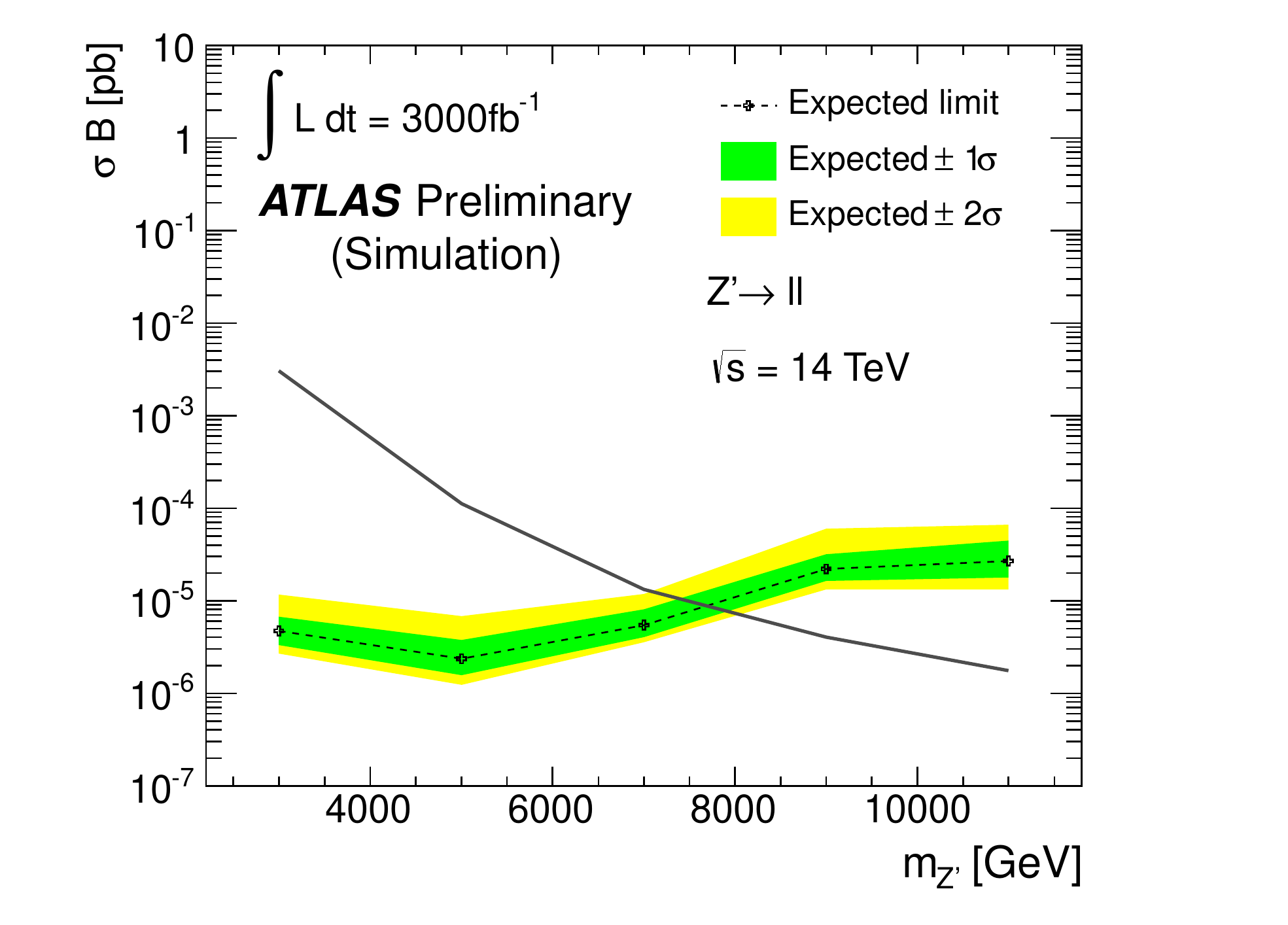}
  \caption{\label{fig:zprimemu} Left: The reconstructed dimuon mass
  spectrum for the $Z^\prime$ search with $3000\,\ifb$ of $pp$ collisions at $\sqrt{s} =14 \TeV$. The highest-mass bin includes the overflow. Right: he 95\% CL upper limit on the cross section times branching ratio. Also shown is the theoretical expectation for the $Z^\prime_{SSM}$ cross section.}
\end{figure}

Templates of the $m_{\ell\ell}$ spectrum are constructed for the
background plus varying amounts of signal at different resonance
masses and cross sections.
The Sequential Standard Model (SSM) $Z^\prime_{SSM}$ boson, which has
the same fermionic couplings as the Standard Model $Z$ boson, is used
as the signal template.
The $m_{\ell\ell}$ distribution, for events above 200 GeV, and the resulting limits as a function
of $Z^\prime_{SSM}$ pole mass are shown in Figs.~\ref{fig:zprimeel}
and~\ref{fig:zprimemu} for the $ee$ and $\mu\mu$ channels, respectively.

The 95\% CL expected limits in the absence of signal, using statistical errors only, are shown
in Table~\ref{tab:zprime}. The increase of a factor of ten, from 300 to 3000 fb$^{-1}$ in integrated luminosity raises the sensitivity to high-mass dilepton resonances by up to 1.3 TeV.

\begin{table}[h!]
\centering
\begin{tabular}{lccc}
\hline
model              & $300\,\ifb$  & $1000\,\ifb$     & $3000\,\ifb$  \\
\hline
\hline
$Z^\prime_{SSM} \to ee$ & 6.5  & 7.2   & 7.8   \\
$Z^\prime_{SSM} \to \mu \mu$ & 6.4  & 7.1   & 7.6   \\
\hline
\end{tabular}
  \caption{Summary of the expected limits for
      $Z^\prime_{SSM} \to ee$ and $Z^\prime_{SSM} \to \mu \mu$
      searches in the Sequential Standard Model for $pp$ collisions
      at $\sqrt{s} = 14 \TeV$.
      All limits are quoted in TeV.}
\label{tab:zprime}
\end{table}

%% file: top.tex
\section{Flavor-Changing-Neutral-Currents in Top-Quark Decays}
\label{sec:fcnc}
 Within the Standard Model,
flavor-changing-neutral-current (FCNC) decays are forbidden at tree
level due to the GIM mechanism~\cite{Glashow:1970gm}, and highly
suppressed at loop level with branching fractions below
$10^{-12}$~\cite{DiazCruz:1989ub,Eilam:1990zc,Mele:1998ag,AguilarSaavedra:2002ns},
which are inaccessible even at HL-LHC. Therefore {\em any}
observation of top quark FCNC decays would be a definite indication
of new physics. FCNC decays have been sensitively searched for in
lighter quarks, placing strong constraints on many models of BSM
physics.  Tests of FCNC in the top sector have only recently become
sensitive enough to probe interesting BSM phase space in which the
FCNC branching fraction can be significantly
enhanced~\cite{AguilarSaavedra:2004wm}. Examples of BSM models with
enhanced FCNC top decay rates are quark-singlet (QS) models,
two-Higgs doublet (2HDM) and flavor-conserving two-Higgs doublet (FC
2HDM) models, the minimal supersymmetric model (MSSM), SUSY models
with R-parity violation ($\slashed{R}$), the topcolor assisted
technicolor model (TC2)~\cite{Lu:2003yr}, and models with warped
extra dimensions (RC)~\cite{Agashe:2006wa}. FCNC decay are sought
through $t\rightarrow q\gamma$ and $t\rightarrow qZ$ channels where
$q$ is either an up or a charm quark. Table~\ref{tab:fcnc} shows the
Standard Model and BSM decay rates in the various channels.
The best current direct search limits are 3.2\% for $t\rightarrow \gamma q$~\cite{Abe:1997fz} and 0.21\% for $t\rightarrow Zq$~\cite{Chatrchyan:2012hqa}.
\begin{table}
\begin{tabular}{|c|c|c|c|c|c|c|c|c|}
  \hline
  Process & SM & QS & 2HDM & FC 2HDM & MSSM & $\slashed{R}$ & TC2 & RS \\ \hline\hline
  $t\rightarrow u\gamma$ & $3.7\times 10^{-16}$ & $7.5\times 10^{-9}$ & --- & --- & $2\times 10^{-6}$ & $1\times 10^{-6}$ & --- & $\sim 10^{-11}$ \\
  $t\rightarrow uZ$ & $8.0\times 10^{-17}$ & $1.1\times 10^{-4}$ & --- & --- & $2\times 10^{-6}$& $3\times 10^{-5}$ & --- & $\sim 10^{-9}$ \\
  $t\rightarrow ug$ & $3.7\times 10^{-14}$ & $1.5\times 10^{-7}$ & --- & --- & $8\times 10^{-5}$ & $2\times 10^{-4}$ & --- & $\sim 10^{-11}$ \\\hline\hline
  $t\rightarrow c\gamma$ & $4.6\times 10^{-14}$ & $7.5\times 10^{-9}$ & $\sim 10^{-6}$ & $\sim 10^{-9}$ & $2\times 10^{-6}$ & $1\times 10^{-6}$ & $\sim 10^{-6}$ & $\sim 10^{-9}$ \\
  $t\rightarrow cZ$ & $1.0\times 10^{-14}$ & $1.1\times 10^{-4}$ & $\sim 10^{-7}$ & $\sim 10^{-10}$ & $2\times 10^{-6}$ & $3\times 10^{-5}$ & $\sim 10^{-4}$ & $\sim 10^{-5}$ \\
  $t\rightarrow cg$ & $4.6\times 10^{-12}$ & $1.5\times 10^{-7}$ & $\sim 10^{-4}$ & $\sim 10^{-8}$ & $8.5\times 10^{-5}$ & $2\times 10^{-4}$ & $\sim 10^{-4}$ & $\sim 10^{-9}$ \\
  \hline\hline
\end{tabular}
\caption{Branching fractions for top FCNC decays for the Standard Model and BSM extensions.  References are given in the text.}
\label{tab:fcnc}
\end{table}
A model-independent approach to top quark FCNC decays using an effective Lagrangian~\cite{top38, top39, Beneke:2000hk} is used here to evaluate the sensitivity of ATLAS in the HL-LHC era. Even if the LHC does not measure the top quark FCNC branching ratios, it can test some of these models or constrain their parameter space, and improve significantly the current experimental limits
on the FCNC branching ratios.

Top quark FCNC decays are sought in top quark pair production in
which one top (or anti-top) decays to the SM $Wb$ final state, while
the other undergoes a FCNC decay to $Zq$ or $\gamma q$. The
sensitivity is evaluated selecting events as in ~\cite{Aad:2012ij}
for the $t\rightarrow Zq$ channel and ~\cite{Aad:2009wy} for the
$t\rightarrow \gamma q$ channel. For the $t\rightarrow \gamma q$
channel, the dominant backgrounds are $t\bar{t}$, $Z+$jets and
$W+$jets events. For the $t\rightarrow Zq$ channel, the background
is mainly composed of $t\bar {t}$, $Z+$jets and $WZ$ events.

\begin{figure}[htb]
\centering
 \includegraphics[width=.70\textwidth]{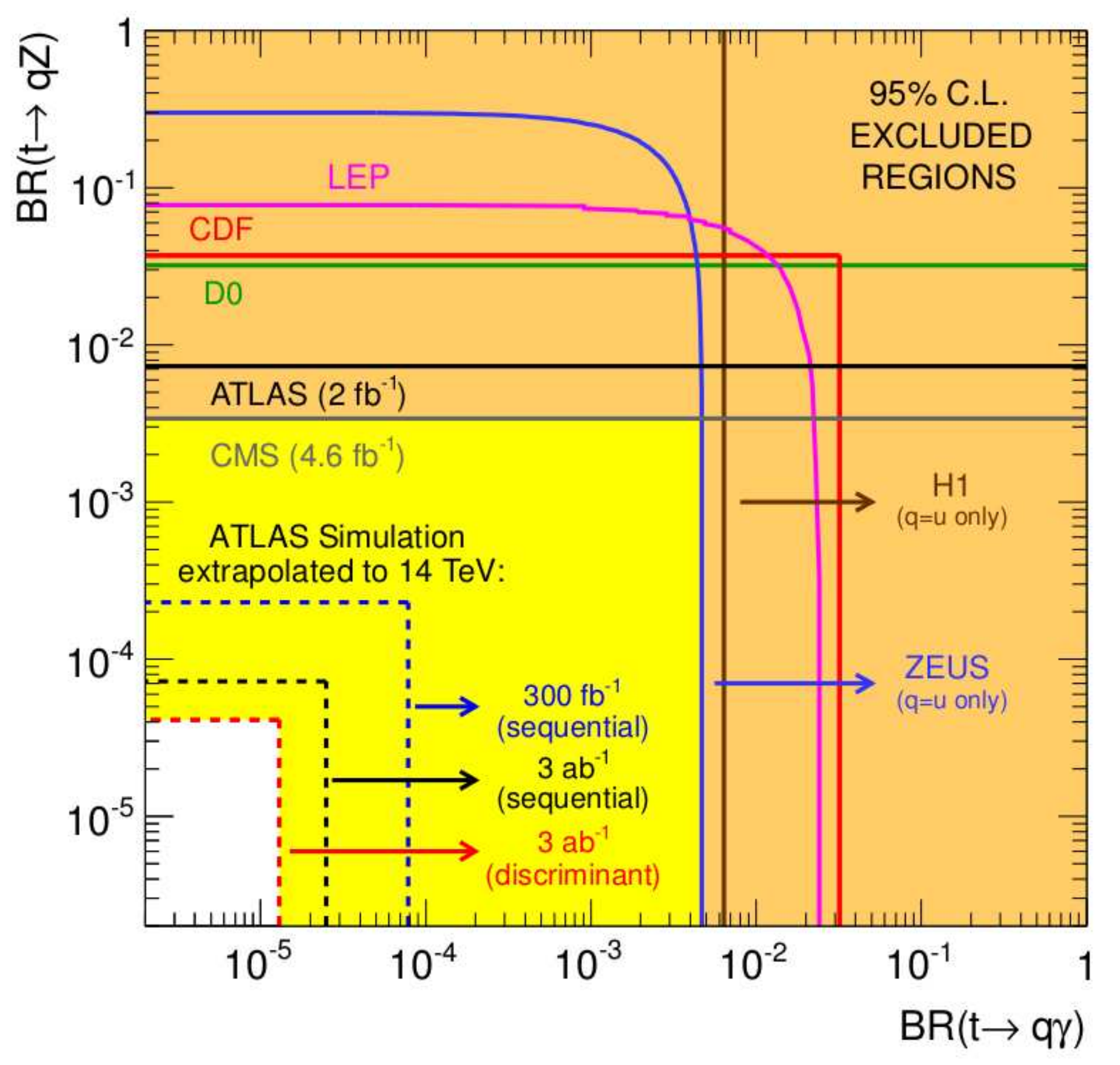}
  \vspace*{-1em}
  \caption{The present 95\% CL observed limits on the $BR({t}\to \gamma q)$
    vs. $BR({t} \to {Zq})$ plane are shown as full lines for the LEP, ZEUS,
    H1, D0, CDF, ATLAS and CMS collaborations. The expected sensitivity at
    ATLAS is also represented by the dashed lines. For an integrated
    luminosity of $L=3000\,\ifb$ the limits range from $1.3\times10^{-5}$ to
    $2.5\times10^{-5}$ ($4.1\times10^{-5}$ to $7.2\times10^{-5}$) for the
    ${t}\to~\gamma q$ (${t}\to~{Zq}$) decay. Limits at $L=300\,\ifb$ are
    also shown.}
  \label{fig:fcncBr}
\end{figure}

In the absence of FCNC decays, limits on production cross-sections
are evaluated and converted to limits on branching ratios using the
SM $t \overline{t}$ cross-section. The HL-LHC expected limits at
95\% CL for the $t\rightarrow \gamma q$ and the $t\rightarrow Zq$
channels, are in the range between $10^{-5}$ and $10^{-4}$.
Figure~\ref{fig:fcncBr} shows the expected sensitivity in the
absence of signal, for the ${t}\to {q}\gamma$ and ${t}\to {qZ}$
channels. Here the lines labeled `sequential' correspond to a sensitivity extrapolated from the analysis done with the 7 TeV data~\cite{Aad:2012ij}.  Those labeled `discriminant' correspond to a dedicated analysis using 14 TeV TopRex Monte Carlo~\cite{Slabospitsky:2002ag} data and a likelihood discriminant. Further improvements could come from the use of more sophisticated analysis discriminants.

%% file: conclusions.tex
\section{Conclusions}
Studies illustrating the physics case of a high-luminosity upgrade of the LHC have
been presented. 
In general, very important gains in the physics reach are
possible with the HL-LHC dataset of $3000\,\ifb$, and some studies are only viable with this high integrated luminosity.  
The precision on the production cross section times branching ratio for most Higgs boson decay modes can be improved by a factor of two to three. 
Furthermore, the rare decay mode of the Higgs boson $H \rightarrow \mu
\mu$ only becomes accessible with $3000\,\ifb$.
When results from both experiments are combined,
first evidence for the Higgs self-coupling may be within reach, representing a fundamental test of the Standard Model.
In searches for
new particles, the mass reach can be increased by up to 50\% with the high-luminosity dataset.

The luminosity upgrade would become even more interesting if new phenomena are
seen during the $300\,\ifb$ phase of the LHC, as the ten-fold
increase in luminosity would give access to measurements of the new physics.

To reach these goals a detector performance similar to that of the present one
is needed, however under much harsher pileup and radiation conditions than today. 
The ATLAS Collaboration is committed to preparing detector upgrades that will realize the potential of HL-LHC operations with the goal of an integrated luminosity of $3000\,\ifb$.